\documentclass[12pt]{article}
\usepackage{epsfig}

\def\beq{\begin{equation}}
\def\eeq{\end{equation}}
\def\bea{\begin{eqnarray}}
\def\eea{\end{eqnarray}}
\def\noi{\noindent}

\begin{document}

\rightline{March 2001}
\rightline{UCOFIS 1/01}
\rightline{US-FT/2-01}
\rightline{LPT Orsay 01-15}
\vspace{0.5cm}

\begin{center}
{\Large\bf Monte Carlo model for nuclear collisions\\
from SPS to LHC energies}\vspace{0.5cm}

N. S. Amelin$^{a,\dag}$,
N. Armesto$^{b,\ddag}$,
C. Pajares$^{c,\S}$
and
D. Sousa$^{d,\P}$

\vspace{0.2cm}

$^a$ {\it Department of Physics,
University of Jyv\"askyl\"a,}\\
{\it P.O.Box 35, FIN-40351 Jyv\"askyl\"a, Finland}\\

$^b$ {\it Departamento de F\'{\i}sica, M\'odulo C2, Planta baja, Campus
de Rabanales,}\\
{\it Universidad de C\'ordoba, E-14071 C\'ordoba, Spain} \\

$^c$ {\it Departamento de F\'{\i}sica de Part\'{\i}culas, Universidade
de Santiago de Compostela,}\\
{\it E-15706 Santiago de Compostela, Spain}\\

$^d$ {\it Laboratoire de Physique Th\'eorique, Universit\'e de Paris
XI,}\\
{\it B$\hat{a}$timent 210, F-91405 Orsay Cedex, France}\\

\end{center}

\noi{\footnotesize $^\dag$ E-mail: nikolai.ameline@phys.jyu.fi
\hskip 3cm $^\ddag$ E-mail: fa1arpen@uco.es}

\noi{\footnotesize $^\S$ E-mail: pajares@fpaxp1.usc.es
\hskip 3.72cm $^\P$ E-mail: Dolores.Sousa@th.u-psud.fr}

\vfill
{\small
\centerline{\bf Abstract}
A Monte Carlo model to simulate nuclear collisions in the energy range
going from SPS to LHC, is presented.
The model includes in its initial stage both soft and
semihard components, which lead to the formation of color strings.
Collectivity is taken into account considering the
possibility of strings in color representations higher than triplet
or antitriplet, by means of string fusion. String breaking leads to
the production of secondaries. At this point, the model can be used
as initial condition for further evolution by
a transport model. In order to tune the
parameters and see the results in nucleus-nucleus collisions, a naif
model for rescattering of secondaries is introduced.
Results of the model
are compared with experimental data, and predictions for RHIC and
LHC are shown.
}

\newpage
\section{Introduction}
\label{intro}

With the announcement of the discovery of Quark Gluon Plasma (QGP) at
the Super Proton
Synchrotron (SPS) at CERN
\cite{qgpan}, the experimental heavy ion program moves now to the higher
energies of the Relativistic Heavy Ion Collider (RHIC) at Brookhaven and
the Large Hadron Collider (LHC) at CERN. Whether this claim can be
considered conclusive or not (see e.g. \cite{gyulassy}), the
most compelling experimental findings at the SPS
\cite{na50,wa97,na49,reanalysis,phi1,phi2,dilep}
are interpreted as positive signatures of QGP
only when conventional, non-QGP models fail to reproduce them. Therefore,
even in the case that QGP has already been obtained, it
is most important that conventional models employed at the SPS become
generalized for RHIC and LHC: They can be used to
describe collisions between less
massive nuclei or more peripheral events than those in which QGP is
expected, and to establish the background to events with QGP production.

On the other hand, the situation with conventional models is not clear
at all. The description of a high energy collision between heavy ions is
a complex task which involves different physical aspects. Predictions
from different models for RHIC and LHC are far from being compatible,
see the reviews \cite{lastcall} and \cite{noso}. For example, the values
for central rapidity densities of charged particles coming from
different models lie in the ranges $600\div 1500$ for central AuAu
collisions at RHIC and $2000\div 8000$ for central PbPb collisions at
LHC.

In this paper a non-QGP model for collisions between nucleons or
nuclei in the energy range going from SPS energies ($\sim 20$
GeV per nucleon
in the center of mass) to LHC energies (5.5 TeV per
nucleon in the center of mass)
is presented (different steps in this direction can be found in
\cite{sfm,asgabp}). The model is based on the ideas of Dual Parton Model
(DPM) \cite{dpm} or Quark-Gluon String Model (QGSM) \cite{qgsm},
considering both soft
and
semihard components on a partonic level. These elementary partonic
collisions lead to the formation of color strings.
Collectivity is taken into account considering the
possibility of strings in color representations higher than triplet
or antitriplet, by means of string fusion, as done in \cite{sfm,rqmd}
(see related approaches in \cite{urqmd,vance}).
String breaking leads to
the production of secondaries.
In this form, the model can be used as initial condition for subsequent
evolution using a transport model, as those of \cite{rqmd,urqmd}.
Nevertheless, in order to tune the
parameters of the model and apply it to nucleus-nucleus collisions,
rescattering between secondaries is
considered on the basis of $2\longrightarrow 2$ collisions, using a very
simple model which allows us just to estimate the effects of such
process. The results
of the code turn out to agree reasonably with existing experimental data
on total multiplicities, and longitudinal and transverse momentum distributions,
and semiquantitatively with strangeness production and stopping power.

The paper is organized as follows: In Section \ref{initial} string
formation will be discussed, both for soft and semihard components,
whose separation will be established. Also in this Section collectivity,
considered as string interaction or fusion, will be presented.
Hadronization of the produced strings will be formulated in Section
\ref{hadro}. In Section \ref{rescatt} our simple approach to
rescattering between secondaries
will be presented. A comparison with experimental data will be done in
Section \ref{comp}, and predictions for RHIC and LHC shown in Section
\ref{pred}, together with some discussion on the
first RHIC data \cite{phobos,phenix}.
In the last Section we will summarize our conclusions and briefly
compare with other approaches.

\section{Initial stage}
\label{initial}

\subsection{Elementary partonic collisions}

To compute the number of elementary partonic collisions we have to
generate the partonic wave functions of the colliding hadrons.
The steps to generate this wave function for the projectile $A$ and
target $B$
are the following: First, the impact parameter $b$ of the collision is
generated uniformly between 0 and $R_A+R_B$ (in the case of nucleons, the
total cross section determines the corresponding radius).

Second, the nuclear wave function is computed. Nucleon
positions inside the nucleus
are distributed in transverse space according to a Woods-Saxon
distribution for $A>11$,
\beq
\label{eq1}
\rho (r) \propto \frac{1}{1+ \exp [(r - r_n)/a]}\ ,
\eeq
with $r_n=1.07 A^{1/3}$ fm and $a=0.545$ fm,
and according to a Gaussian distribution for $A \le 11$, with parameters
chosen for each nucleus \cite{nucleus}.
Then, Fermi motion is given to the
nucleons in the nuclei uniformly in the range $0<p<p_F$, with the
maximum Fermi momentum given in the local Thomas-Fermi approximation
\cite{thomas} by
\beq
\label{eq2}
p_F=h\ [3\pi \rho(r)]^{1/3},
\eeq
with $h=0.197$ fm GeV/c.

Now partons are generated inside each nucleon. Its number is given by a
Poisson distribution \cite{abra},
\beq
\label{eq3}
W_N\propto
\exp{(-g(s)C/2)}\ \frac{[Cg(s)/2]^N}{CN!/2}\ ,
\ \ g(s)=g_0s^
{\Delta/2},\ \ g_0=\frac{8\pi \gamma_P}{\sigma_P},
\eeq
with $\Delta=0.139$ the pomeron intercept minus 1, $C=3.0$ the
quasieikonal parameter which takes into account low mass nucleon
dissociation,
$\gamma_P=1.77$ GeV$^2$ the pomeron-nucleon vertex,
$\sigma_P
=3.3$ mb the parton-parton cross section and $\sqrt{s}$ the center of
mass energy for each nucleon-nucleon collision.

Parton positions in transverse
space (inside a nucleon) are given by a Gaussian according to Regge
theory,
\beq
\label{eq4}
F(b)=\frac{1}{4\pi\lambda}\ \exp{\left(-\frac{b^2}{4
\lambda}\right)},\ \ \lambda=R_0^2+\alpha^\prime \ln{s},
\eeq
with $R_0^2=3.18$ GeV$^2$ and $\alpha^\prime=0.21$ GeV$^2$ the pomeron
slope.

Now, one parton from the projectile and one from the
target produce an inelastic collision if both are
within an area in impact parameter equal to
$\sigma_P=2\pi r_P^2$,
$r_P=0.23$ fm.
In this way, events with no inelastic collisions are elastic, while
those with at least one inelastic collision are inelastic. Taking
the total cross section given by the quasieikonal model \cite{quasiei}
\beq
\sigma_{tot}=
\sigma_P\ f(z/2),\ \ 
z=
\frac{C\gamma_P s^\Delta}{\lambda}\ 
,\ \ f(z)=\sum_{k=1}^\infty \frac{(-z)^{k-1}}{k\cdot k!}\ ,
\label{eq5}
\eeq
all cross sections can be computed, see next Subsection
(all formulae reduce to the usual
eikonal ones with $C=2$).

\subsection{Semihard component}

The inclusion of semihard components, in the form of a two-component
model, is needed to  reproduce the $p_T$
spectra in hadronic collisions, see Section \ref{comp}. In the model
this is performed considering that an inelastic collision is hard with
probability
\begin{equation}
W_h=\frac{C_h\ (s-s_0)^{\Delta_h}}{C_h\
(s-s_0)^{\Delta_h}+s^{\Delta}}\ ,
\label{eq6}
\end{equation}
with $\Delta_h=0.50$, $\sqrt{s_0}=25$ GeV and $C_h=0.0035$. A hard collision
proceeds through the packages PYTHIA 5.5 + ARIADNE 4.02 + JETSET 7.3
\cite{pyth,ariad}. Only gluon-gluon collisions are included in PYTHIA,
and the key parameter here is the cut-off in transverse momentum
\beq
p_{T min}=3.03+0.11\ln{(s/s_0)}\ \  
{\rm GeV/c}. \label{eq7}
\eeq

The minimum energy for an elementary collision
to be accepted by PYTHIA is 20 GeV, and for the global collision the
minimum center of mass energy per nucleon is
$\sqrt{s_0}=25$ GeV. An event is considered hard if at least one of its
inelastic elementary collisions successfully proceeds through PYTHIA.

While the concrete choice of the parameters in $p_{T min}$ comes from
a fit to experimental data, let us make some comments on its functional
form.  In our case,
an increase of $p_{T min}$ with increasing energy makes possible a smooth
transition from the soft to the semihard part of the $p_T$ spectrum.
Usually $p_{T min}$ is taken as either constant or increasing as a
polynomial of
a logarithm of $s$
\cite{hijing,dpmjet}. It may be argued that the $p_{T min}$ value which
indicates the transition from nonperturbative to perturbative QCD (pQCD), is
related with the proposed saturation scale $Q_s^2$ \cite{satur1,satur2}: below
this $Q_s^2$, the number of partons in the hadron wave function cannot
grow, as new partons fuse with the existing ones and cannot
be resolved individually. Nevertheless, apart from conceptual
differences, the dependences of
$p_{T min}$ and $Q_s^2$ are not the same:
while the first depends only on energy, the
second one also depends on the size of the colliding objects ($Q_s^2
\propto A^\alpha$, $\alpha = 1/3 \div 2/3$).

\begin{figure}[htb]
\begin{center}
\epsfig{file=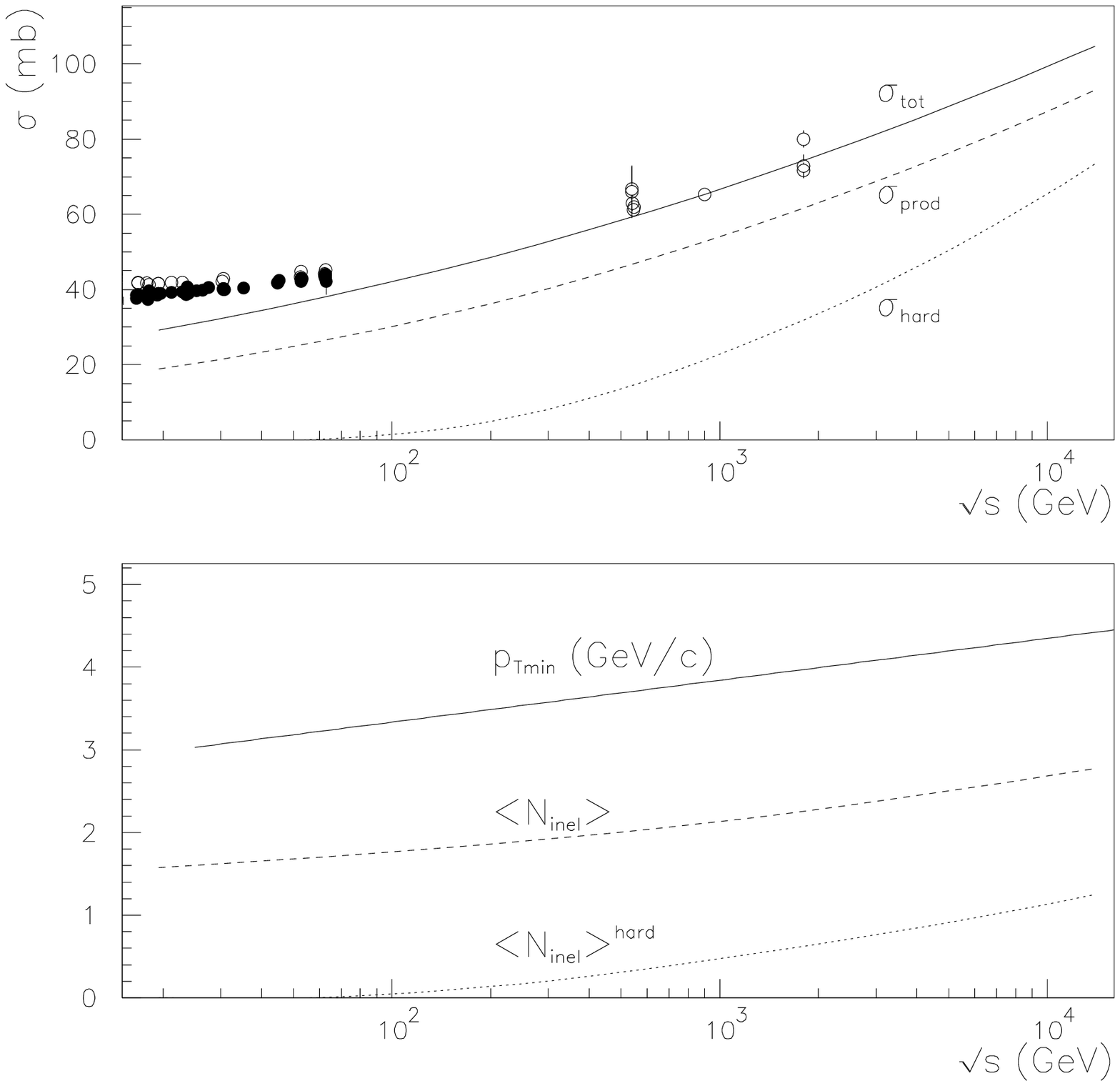,width=15.5cm}
\end{center}
\vskip -1.0cm
\caption{}
\label{fig1}
\end{figure}

Results of the model for the total, inelastic (production) and hard
cross sections in pp and $\bar {\rm p}$p collisions at different energies
are shown in Fig. \ref{fig1} and compared with experimental
data for the total cross section \cite{pdg}.
It can be observed that both the total
and the production cross section are too small at low energies, while
they get reasonable values at higher energies. The reasons for the
existing discrepancies are three: In first place,
diffraction is not properly included in
the model, so it is difficult to distinguish between production and
inelastic cross sections. In second place, no reggeon contribution
(decreasing with energy) has been included.
In third place, at the level of the cross
sections no distinction is made between nucleons and antinucleons as
projectiles and targets. These two last reasons should improve the
agreement with data at energies of SPS and Intersecting Storage Rings (ISR).
Also in this Figure it is
shown the value of $p_{T min}$ and the mean number of total and hard inelastic
collisions per event.

\subsection{String formation and fusion}
\label{strfus}

Each soft parton-parton collision gives rise to two strings
\cite{dpm,qgsm}, stretched
either between valence quarks and
diquarks (for the first collision suffered by a nucleon) or sea
quarks and antiquarks (for the subsequent ones). For the latter, their
flavors follow the ratio
$u:d:s=1:1:0.26$. Hard collisions proceed through PYTHIA as $gg
\longrightarrow gg$. For the string ends and hard gluons, the
longitudinal momentum fractions are distributed as
\beq
L(x_1,x_2,\dots ,x_n)=
f_{qq}(x_1)f_q(x_2)\cdots f_q(x_n),
\ \ \sum_{k=1}^n x_k
=1. \label{eq8}
\eeq

For soft strings ends, the individual momentum distributions are those
of the QGSM \cite{qgsm},
\begin{equation}
f_{qq}(x)=x^{3/2},\ \  f_{q(\bar q)}(x)=\frac{1}{\sqrt{x}}\ ,
\label{eq9}
\end{equation}
with a lower cut-off $x_{min}=0.3\ {\rm GeV}/\sqrt{s_{NN}}$ to ensure
that the strings have mass enough to be projected onto hadrons,
$\sqrt{s_{NN}}$ being the center of mass energy per nucleon.

For partons involved in hard collisions, the longitudinal momentum
fractions are taken by PYTHIA from PDFLIB \cite{pdflib}, with the
possibility of considering the difference of parton
distributions inside nuclei given by the parametrization EKS98
\cite{eks} or by a parametrization as $F_{2A}$ \cite{eqc}. After
generating the final gluons, each of them splits into a $(q\bar q)$ pair
and strings are stretched between them, according with the standard
procedure in PYTHIA \cite{pyth}.

The transverse momentum of both partons at the string ends and hard
partons, coming from a nucleon which has been wounded $m$ times,
is given by a Gaussian:
\begin{equation}
T(p_\perp)=\frac{1}{\pi \delta^2} \
\exp{(-p_\perp^2/\delta^2)},\ \ \delta=0.5\ \sqrt{m} \ \ {\rm GeV/c};
\label{eq10}
\end{equation}
in this way, $p_T$-broadening is taken into account \cite{cfk}.

The number of strings exchanged in one collision is quite low for
nucleon-nucleon collisions, but this number increases with energy, size
of projectile and target and centrality of the collisions. Strings can
be viewed as objects with a certain area, given by the uncertainty
relation as $\propto 1/\langle p^2_T \rangle$, in the transverse plane of the
collision. When the
number of strings is high enough, they begin to overlap and the usual
hypothesis in QGSM or DPM of the strings being independent sources of
secondary particles is expected to break down. A possible way of
considering this is to compute the density of strings in the transverse
plane and use two-dimensional percolation as an indicator of the onset
of collectivity \cite{percol1,percol2}. Percolation takes place when
domains of overlapping strings acquire a size of the order of the total
available size for the collision.

While percolation is a second order phase transition, the option we use
in this model, fusion of strings, does not lead to any phase transition
\cite{bpr}. In the model, ordinary strings (i.e. in a triplet representation of
SU(3)) fuse\footnote{A similar mechanism exists in RQMD
\protect{\cite{rqmd}}, called color ropes.}
in pairs when their parent partons (those which determined
the inelastic collision the strings come from) are within a certain area
$\sigma_{fus}=2\pi r^2_{fus}$
in impact parameter space.
In the code we consider
only fusion of two strings but there is a probability of fusion of more
than two. An
effective way of taking this into account is to increase the cross
section for the fusion of
two strings, for which  we will take $\sigma_P<\sigma_{fus} = 7.5$ mb
($r_{fus}=0.35$ fm). This value
is crucial to
reproduce the strangeness enhancement
in central SS and SAg collisions
at SPS \cite{sfm2}. The fusion can take
place only when the rapidity intervals of the strings overlap. It is
formally described by
allowing partons to interact several times, the number of interactions
being the same
both for projectile and target.

The quantum numbers of the fused strings are determined by the
interacting partons and their energy-momentum is the sum of the
energy-momenta of their
ancestor strings. The color charge
of the resulting string ends is obtained
according to the SU(3) composition laws:
\beq
\{3\}\otimes \{3\}=\{6\}\oplus \{\bar 3\},\ \ \{3\}\otimes
\{\bar 3\}=\{1\}\oplus \{8\}.
\label{eq11}
\eeq
Thus, two triplet strings fuse into either a sextet or an antitriplet
string with probabilities 2/3 and 1/3 respectively, and one triplet and
one antitriplet string fuse into either a singlet or an octet string
with
probabilities 1/9 and 8/9 respectively.

Two comments are in order: On the one hand and as written above, the
fusion of strings means nothing
related to a phase transition. On the contrary, percolation of strings
\cite{percol1} is
a non-thermal second order phase transition. In this case, the key
parameter is
$\eta = \pi r^{2} N / (\pi R_A^{2})$, which is the density of strings
$N/(\pi
R_A^{2})$ (number of
strings $N$ produced in the overlapping area of the collision, $\pi
R_A^{2}$ for central collisions) times the
transverse size of one string $\pi r^{2}$. The critical point
for percolation is $\eta_{c} \simeq
1.12 \div 1.5$ depending on the profile function of
the colliding nuclei \cite{percol2}. With $r \simeq 0.2\div 0.25$ fm,
this critical value means $6 \div 12$ strings/fm$^{2}$. The value of 9
is reached
in central PbPb collisions at SPS, in central AgAg collisions at RHIC
and in central
SS collisions at LHC. We expect for $\eta$ around or greater than
$\eta_{c}$, that
the approximation of fusion of just two strings fails.
On the other hand, only fusion of soft strings is considered. Hard
strings are not fused, their area being proportional to
$1/p_{T}^{2}$. Some effect of the fusion of such strings could appear
at LHC energies where, for instance, in central PbPb collisions they
amount for 32 \% of the
binary nucleon-nucleon collisions.

\section{Hadronization}
\label{hadro}

Now we consider the breaking of a soft string with color charges $Q$ and
$\bar Q$ in its ends (corresponding to a representation $\{N\}$ of
SU(3)).
In our model, it is due to the production of two (anti)quark complexes
with the same
color charges $Q$ and $\bar Q$ as those at the
ends of the string \cite{sfm}\footnote{This possibility is the dominant
one
for strings formed by fusion of two triplet strings \protect{\cite{sfm}}.
For higher color representations, production of quark complexes with
color charges $Q^\prime<Q$ begins to dominate. This option is taken in
RQMD \protect{\cite{rqmd}}; nevertheless, the close similarities in the
consequences of string fusion in both approaches, strongly suggest that
the difference in the breaking mechanism can be compensated by a
different choice of the fragmentation parameters.}.
The probability rate is given by the
Schwinger formula \cite{schwin}
\beq
W \propto K^{2}_{\{N\}} \exp{(-\pi M_T^{2} / K_{\{N\}})},
\label{eq12}
\eeq
where $K_{\{N\}}$ is the string tension for the $\{N\}$
representation,
proportional to the corresponding quadratic Casimir operator
$C^{2}_{\{N\}}$ (as found both in lattice QCD and in the Stochastic
Vacuum Model \cite{bali,simonov}), i.e.
\beq
K_{\{N\}}=K_{\{3\}} {C^{2}_{\{N\}} \over C^{2}_{\{3\}}}\ ,\ \
C^{2}_{\{3\}}=
4/3,\ \ C^{2}_{\{6\}}=10/3,\ \
C^{2}_{\{8\}}=3.
\label{eq13}
\eeq

For the longitudinal breaking of the string, an invariant area law
\cite{artru} is employed,
\beq
P\propto \exp{(-bA)},\ \ K_{\{N\}}\propto b C^{2}_{\{N\}},\ \  A=p_+p_-
\label{eq14}
\eeq
being the area in light-cone
momentum space determined by the breaking point in the
center of mass frame of the string. This law gives results quite similar to
those of the Lund model \cite{lund} implemented in JETSET \cite{pyth}.

We proceed as follows: Eq. (\ref{eq12}) is used to decide the flavors of
the quark and antiquark complexes created. We take $K_{\{3\}}=0.18$
GeV$^2$ and $m_u=m_d=0.23$ GeV/c$^2$, $m_s=0.35$ GeV/c$^2$, and the
masses of a complex $(q_1\dots q_l)$ is given by $M(q_1\dots
q_l)=\sum_{i=1}^l m_{qi}$. Then $p_T$ is given to one of the created
complexes and $-p_T$ to the other one, according to a Gaussian law
\beq
f(p^2_T)\propto \exp{(-\alpha_{\{N\}} p^2_T)},
\label{eq15}
\eeq
with $\alpha_{\{3\}}=\alpha_{\{\bar 3\}}=4$ GeV$^{-2}$ and
\beq
\alpha_{\{N\}}= 2 \alpha_{\{3\}} {C^{2}_{\{3\}} \over C^{2}_{\{N\}}}\ ,\
\ \{N\}\neq \{3\},\{\bar 3\}.
\label{eq16}
\eeq
Finally
a breaking point is sampled according to
Eq. (\ref{eq14}) in the available phase space, with $b=1.83$
GeV$^{-2}$.

Fragmentation proceeds in an iterative way: String fragments are taken
as new strings which are broken again, until the mass of the created
fragments is too low to allow further breaking (i.e. projection onto
hadrons with the right quantum numbers). Then these final
fragments (and those fused strings resulting in the singlet $\{1\}$
representation) are treated as quark clusters and decayed according to
combinatorics and phase space. Spin of the produced particles is
constructed according to SU(2) considerations.

The main consequences of string fusion are a strong reduction of
multiplicities (both due to the energy-momentum conservation and to the
reduction of the effective number of sources of secondaries)
and a slight increase of $\langle p_T^2\rangle$
\cite{sfm}, an increase in baryon and strangeness production
\cite{sfm,sfm2}, a strong increase in the cumulative effect \cite{cumu}
and a decrease in forward-backward correlations \cite{foba}.

On the other hand,
strings produced in hard collisions (only $gg\longrightarrow gg$)
are managed by PYTHIA + ARIADNE +
JETSET \cite{pyth,ariad}\footnote{Although it is not relevant for the
generation of momenta of final
gluons, in PYTHIA and JETSET the value of $\Lambda_{QCD}$ has been taken
from the corresponding set of parton densities for 5 flavors.}.
For ARIADNE,
PARA(6) is fixed so as the
transverse momentum of the radiated gluon should be less than that of
the hard gluon (i.e. the one participating
in the $gg$ scattering)
and MSTA(9)=MSTA(14)= MSTA(31)=0.
In JETSET,
PARJ(41)=1.7 GeV$^{-2}$ and PARJ(42)=0.6 GeV$^{-2}$; besides,
PARJ(21)=0.55 GeV/c.
Only production of three flavors (u,d,s)
is considered in the present implementation of the model, so it cannot
be used to study production of heavier flavors (some steps in this
direction were done in \cite{asgabp}).
As a last point, MSTU(4), MSTU(5) and the dimensions in
LUJETS have been set to 120000, which has shown to be enough for central
PbPb collisions at LHC. All the other parameters and options
in these programs have
been set to their default values.

\section{Rescattering of secondaries}
\label{rescatt}

As stated in the Introduction, in this stage the model can be used as an
initial condition for further evolution, using either a hydrodynamical
model or a microscopic transport as RQMD \cite{rqmd}, UrQMD
\cite{urqmd}, HSD \cite{hsd}, ART \cite{art},$\dots$ (see \cite{asgabp}
for a study of evolution of particle and energy densities).
Nevertheless, it is usually assumed that
the enhancement of hyperons, antihyperons and $\phi$'s observed in
heavy ion experiments at SPS \cite{wa97,na49,reanalysis,phi1,phi2} cannot be
fully explained by using exclusively a mechanism which goes beyond the
independent string hypothesis, as string fusion
\cite{sfm,rqmd,urqmd,vance,sfm2} or baryon junction migration
\cite{vance,dpmjet,dpmdb,bj1,bj2}.
In order to reproduce these experimental features
rescattering of particles in the hadron gas (produced particles among
themselves
and with spectator nucleons) \cite{rescat} has been introduced in many
models.
To tune the code and study
nucleus-nucleus collisions, we will make a very
simple rescattering model with no space-time evolution, fitted to SPS data.
Results of this approach will be presented, but one must keep in mind
that predictions which depend critically on rescattering effects should
be taken with much care.

Our implementation of rescattering
is extremely naif,
trying not to solve the full Boltzmann transport equation for all
particles but only to make a model, as simple as possible, which gives
us an estimation of which effects such rescattering would produce.
Neither formation time nor space-time evolution of secondaries are
properly considered; instead
we require a common minimum density of particles in the rapidity bin of the
considered particles, for rescattering to occur. This minimum density,
$dN/(dydp_T)|_{min}=17$, has
been chosen for rescattering not to affect results in nucleon-nucleon
collisions up to the highest energies.
Rapidity and $p_T$ distances between particles have to be lower than 1.5
units and 0.3 GeV/c respectively. Only two body reactions have been
included, with inverse reactions as required by detailed balance. Spin
is ignored, and rescattering takes place before resonance decay.
All cross sections
are taken equal for all reactions (except for $\Omega$ production
and nucleon annihilation).

Operationally, both products of string breaking and spectators are
randomly ordered into an array $(1,\dots,N)$. We compute the possibility
of rescattering
of the first element with all the others in pairs: $(1,2)$, $(1,3)$,
$(1,4)$,$\dots$
If either in one of the pairs
$(1,j)$, $j=2,\dots,N$, rescattering (either elastic or
inelastic) occurs or we reach pair $(1,N)$ with no kind of
scattering happening, we go to element 2 and examine the pairs $(2,1)$,
$(2,3)$, $(2,4)$,$\dots$
This is repeated until the pair $(N,N-1)$ is
examined. As particles produced in rescattering of the pair $(i,j)$,
$i,j=1,\dots,N$, $i\neq j$, occupy
the same places $i,j$ in the array as their ancestors,
particles produced by rescattering have a
chance to rescatter again.

The probability for two particles to scatter in a given inelastic channel 
is 7 \% (except for channels 
involving $\Omega$'s and nucleon annihilation,
where it is 14 and 70 \% respectively).
For a given process, the probability for elastic scattering is given by
the sum of those corresponding to all inelastic
channels considered for these initial particles. Cross sections
(probabilities) are considered energy independent, except for the
trivial kinematical thresholds, and isotropic in the center of mass of
the colliding secondaries and/or spectators.
The considered reactions (together with those for the corresponding
antiparticles) \cite{rescat} can be classified into:

\begin{itemize}

\item Light pair, $(q\bar q)$,
annihilation to create another light pair,
or light quark exchange:
$\pi N \to \pi N$, $\pi \pi \to \pi \pi$, $\pi Y \to \pi Y$,
$\pi \Xi \to \pi \Xi$,
$K N \to K N$ and $ K Y \to K Y$, where $Y = \Sigma, \Lambda$.

\item Other considered reactions are:
$\pi N \to K Y$, $\pi \pi \to K \bar K$,
$\pi Y \to K \Xi$,
$\pi \Xi \to K \Omega$ and $\bar K N \to \phi Y$. These reactions 
can be classified into:

\begin{enumerate}
\item Light pair, $(q\bar q)$, annihilation to create a $(s\bar s)$ pair.

\item Reactions with baryon exchange (that is, with three lines in the
t-channel).
\end{enumerate}

\item Reactions with strangeness exchange: 
$\bar K N \to \pi Y$, $\bar K Y \to \pi \Xi$, $\bar K \Xi \to \pi \Omega$,
$K Y \to \phi N$, $K \Xi \to \phi Y$ and $K \Omega \to \phi Y$.	
This type of processes can produce
(anti)baryons with
several strange (anti)quarks and are exothermic.

\item Nucleon-antinucleon annihilation into two pions: $N \bar N \to \pi \pi$.
This type of reaction has a much larger
cross section at low energies than reactions
consider before; for this reason its probability has been chosen ten
times larger than the others. This is also an effective way to take 
into account final states involving more than two pions.
  
\end{itemize}

To simplify, particles produced in rescattering are always projected
onto the
lowest spin state.
Decay of resonances proceed through the usual JETSET routines, with
MSTJ(22)=2, and decay of $\pi^0$'s is forbidden.
The results of our rescattering model on strangeness and baryon/antibaryon
production
can be summarized
in three points: hyperon and $\phi$ enhancement, antinucleon
annihilation and a slight increase of stopping power (kinematical
effects of our rescattering model are very small, due to the applied cuts in
rapidity and transverse momentum). Besides, a
slight decrease of
multiplicities appears, as we will see in the next Section.

\section{Comparison with experimental data}
\label{comp}

In order to show the quality of the choice of the parameters, in this
Section we will compare the results of the code with experimental data.
We will also analyze the influence of the different physical mechanisms
implemented in the model.
From now on and
unless otherwise stated, results of the code come from its default
version with string fusion, rescattering (which do not affect results in
nucleon-nucleon collisions, and in pA collisions at SPS energies),
and GRV 94 LO \cite{grv94} parton densities
with EKS98 \cite{eks} nuclear corrections.

\subsection{Hadron-hadron collisions}

Results of the model for
the mean numbers of produced particles in minimum bias pp collisions at
$\sqrt{s}=19.4$ and 27.5 GeV are shown in Tables \ref{tab1} and
\ref{tab2} respectively, compared with experimental data.
An overall agreement can be observed. Two comments are in order: On the
one hand, the influence of fusion in nucleon-nucleon collisions is tiny,
apart from a slight increase in antibaryons. On the other hand, the
number of both $\Lambda$'s and $\bar \Lambda$'s is overestimated in the
model. This is due to the fact that in the model, threshold effects,
important at these low energies, are treated very roughly (see further
comments in the next Subsection). At higher
energies the situation improves. For example, in $\bar {\rm p}$p
collisions at $\sqrt{s}=200$ GeV, the mean number of $\Lambda+\bar
\Lambda$ in the model is 0.56,
to compare with the experimental result $0.46\pm 0.12$
\cite{lambdaua5}.

\begin{figure}[htb]
\begin{center}
\epsfig{file=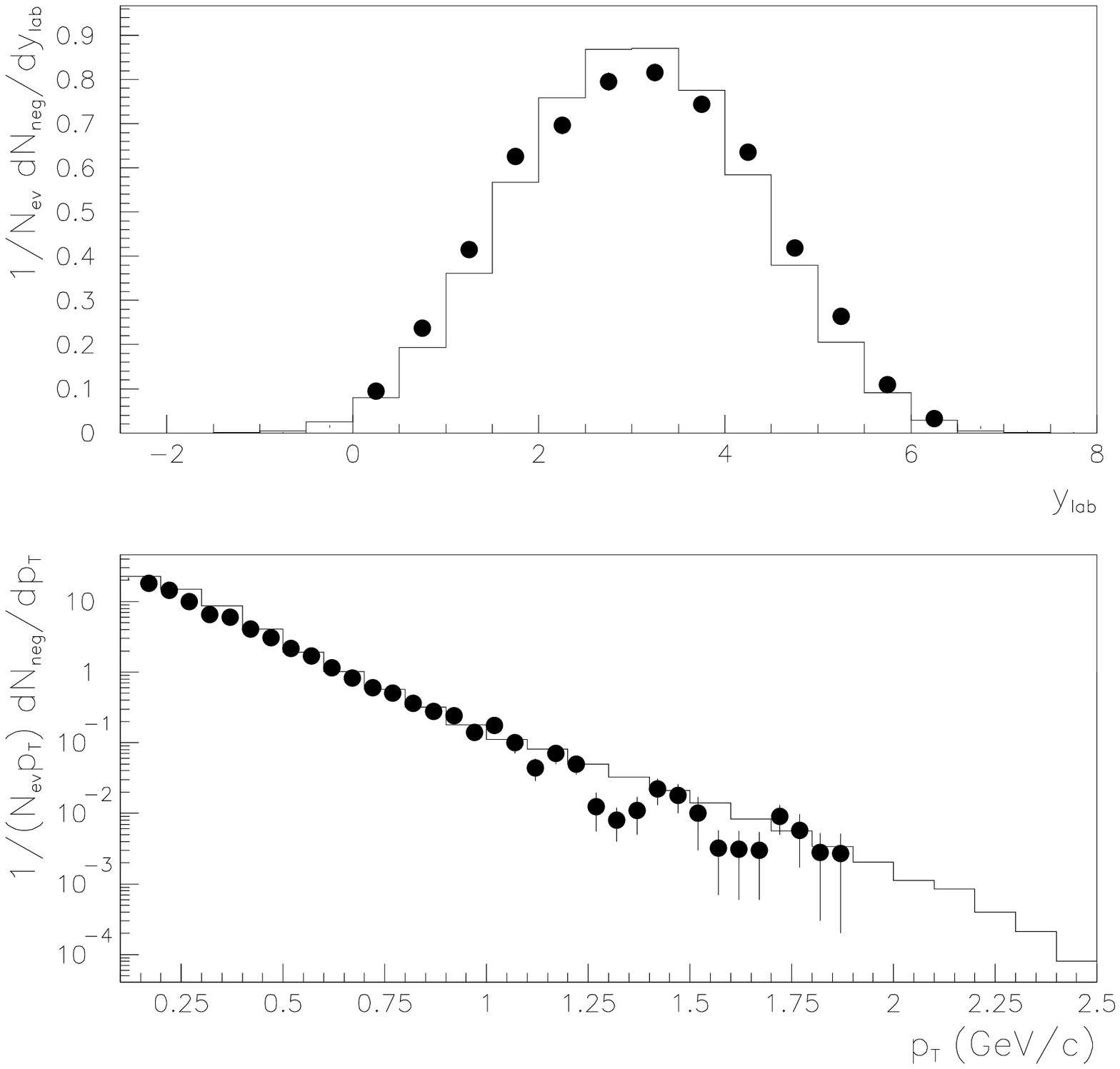,width=15.5cm}
\end{center}
\vskip -1.0cm
\caption{}
\label{fig2}
\end{figure}

In Fig. \ref{fig2}, rapidity and transverse momentum distributions of
negative particles in minimum bias pp collisions at
$\sqrt{s}=19.4$ GeV are shown and compared with experimental data. The
result is satisfactory. In Figs. \ref{fig3} and \ref{fig4}
pseudorapidity and transverse momentum distributions of charged particles
in $\bar {\rm p}$p
collisions at $\sqrt{s}=200$ and 1800 GeV
are compared with experimental data.
The agreement is reasonable,
although the multiplicity at 200 GeV seems to be slightly underestimated.
The results of the model without semihard component are also shown in these
Figs., and they do not describe the $p_T$ distributions, which justifies
the inclusion of hard collisions. Besides, the
results given for different sets of parton distributions, both old
\cite{grv94} and new \cite{cteq5,grv98} and leading order or
next-to-leading order, are very
similar. This fact may look surprising from a pQCD point of view. The
main reason is that the cross sections and the number of inelastic
collisions in our model are determined by Eqs. (\ref{eq3}), (\ref{eq4}),
(\ref{eq5}) and (\ref{eq6}), which are independent of the choice of
partonic distributions in PYTHIA (this is not so in other models, see
e.g. \cite{ranftcomp}). We also
think that the quite high $p_{T min}$ we use in PYTHIA,
Eq. (\ref{eq7}), and the gluon radiation and fragmentation performed by
ARIADNE and JETSET, may have some influence on the fact that no
difference is apparently seen in the transverse momentum distributions.

In Fig. \ref{fig5}, the evolution of the mean transverse momentum of
charged particles is studied in $\bar {\rm p}$p collisions at Sp$\bar
{\rm p}$S, versus the center of mass energy and, for
different particles, versus central charged multiplicity. The trend of
data is reproduced and we find the agreement reasonable (this cannot be
achieved without the hard component, as seen in this Figure).
In Fig. \ref{fig6}
the topological cross section for charged particles in the central
region is examined at different energies for $\bar {\rm p}$p collisions
at Sp$\bar
{\rm p}$S, and the agreement is also reasonable, considering that the
model slightly underestimates multiplicities at 200 GeV but correctly
reproduces those at 1.8 TeV, see Figs. \ref{fig3} and \ref{fig4}.

\subsection{Proton-nucleus and nucleus-nucleus collisions}

In Table \ref{tab3} results of the model in pA collisions are compared with
experimental data on negative multiplicities. An overall agreement is obtained.
The reduction of multiplicities due to string fusion can be observed.

In Table
\ref{tab4}, mean numbers of produced particles are compared with experimental
data, for central SS collisions at SPS energies. The agreement is reasonable.
Only the number of both $\Lambda$'s and $\bar \Lambda$'s in the model is
significantly below
the experimental data. The number of $\Lambda$'s is increased by both
string fusion and rescattering,
while that of $\bar \Lambda$'s is mainly determined by only
string fusion
(see results in PbPb below). Anyhow, rescattering is seen to have little effect in SS.

Let us now discuss PbPb collisions at SPS.
In the last year a large excitement has arisen in the 
heavy ion physics community, related to
the possibility of Quark Gluon Plasma (QGP) 
already been obtained
at SPS energies \cite{qgpan}.
In particular
several signals were mentioned, which point out to the existence of
QGP. Putting aside the abnormal $J/\psi$ suppression and the excess of dileptons
found, there are
three signals related to baryon and strangeness production, namely the
large enhancement
of the (anti)hyperon yields ($\Lambda$, $\Xi$, $\Omega$) in PbPb
collisions compared
to pPb, observed by the WA97 \cite{wa97} and the NA49 \cite{na49} 
Collaborations\footnote{A recent reanalysis \cite{reanalysis} of $\Xi$
data done by the NA49 Collaboration gives yields at midrapidity which are
in much closer agreement to the WA97 \cite{wa97} results than the
previous analysis of NA49 \cite{na49}.};
the linear increase of the inverse exponential
slope of the $m_{T}$ distributions ('temperature') in PbPb
collisions with the mass of the observed particle, except
for $\Omega$
\cite{na49,pt}; 
and the different behavior of the temperature between
pp and AA collisions. 
These
characteristics have been interpreted as the existence of an intrinsic
freeze-out temperature and a
collective hydrodynamical flow which is gradually developed:
firstly, for SS collisions, and, in a more clear way, in PbPb collisions.
In this Subsection we will examine some of these points using our model,
together with other interesting aspects as
$\phi$ production \cite{phi1,phi2}, 
different particle ratios \cite{ratios} and stopping power \cite{stop},

In Fig. \ref{fig7} we show our results for $\Omega$, $\Xi$ and
$\Lambda$ yields for pPb,
and central PbPb collisions at SPS with four different centralities,
together with the experimental
data. In order to disentangle the different processes contributing, in 
Fig. \ref{fig8} it is shown the results of the code 
for central ($b \le 3.2$ fm) PbPb collisions without
string fusion and
rescattering, with string fusion, and with string fusion and rescattering.
A reasonable agreement with data for PbPb is obtained,
only the $\Omega$'s are a
40 \%
below the data and we have some excess of $\bar \Lambda$ and
$\bar \Xi^+$, see next paragraph.
Similar results
have been obtained in the Relativistic Quantum Molecular Dynamics
model \cite{rqmd,lastcall}
by a mechanism of color ropes which consider fusion of
strings; also in the Ultra Relativistic Quantum Molecular
Dynamics model \cite{urqmd}
and in the HIJING model \cite{vance}
by using an ad hoc multiplicative factor in the string 
tension. Also
the Dual Parton Model \cite{dpm}, considering the possibility of creation of 
diquark-antidiquark pairs in the nucleon sea,
together with the inclusion of diagrams which take into account baryon 
junction migration \cite{dpmdb,bj1,bj2}, can reproduce the experimental 
data (for $\Omega$'s some rescattering has still to be added).
The string fusion is the main ingredient to obtain
an enhancement of $\bar \Lambda$ production
and also to reproduce
the $\Xi$ data.
However rescattering seems fundamental to get enough $\Omega$'s.

Nevertheless, our results for pPb are higher than the data for
$\bar \Xi^+$
and $\bar \Lambda$; this last feature
looks quite strange, as $\Lambda$ and $\Xi^-$ yields agree
with data, but we overestimate both $\Lambda$ and $\bar \Lambda$
production in pp collisions at this
energy\footnote{In our opinion, the
comparison of
(anti)hyperon nucleus-nucleus data with those in nucleon-nucleon
collisions
should be taken with caution at
SPS,
because at this relatively low energy the nucleon-nucleon value rises sharply
with increasing energy due to
the $t_{min}$-
and delayed threshold effects \cite{delthr},
which usually are not properly implemented in models.},
see Table \ref{tab1}. As
rescattering plays a minor role in minimum bias pPb collisions, this
turns out to be a result of string fusion.
About $\bar \Lambda$, our results are higher than
the WA97 data also in PbPb,
its production being mainly determined by string fusion and hardly
affected by
rescattering. This fact makes that our results for PbPb are really an
extrapolation in the model from the value for $\bar \Lambda$ production in
central SS
collisions by the NA35 Collaboration, which was used to fix the fusion 
cross section
$\sigma_{fus}$ \cite{sfm2} (even so, the model underestimates $\bar
\Lambda$ production in
central SS, see Table \ref{tab4}). So, from the point of view of our model,
there exists either a large $\bar \Lambda$ annihilation or a conflict between
NA35 data for SS and WA97 data for PbPb and pPb.

In Fig. \ref{fig9} we plot the inverse exponential slopes of the $m_{T}$ 
distributions for different particles, together with the
WA97 experimental data\footnote{The fits have been performed in the same
$m_T$ regions as WA97 did \protect{\cite{pt}}.
For statistical reasons, we compare
the slopes in the model for yields integrated over all rapidities, with
experimental data taken in the central rapidity region.}.
A semiquantitative agreement is obtained. In particular it can be seen that the
$\Omega$ slope does not obey the linear increase with increasing mass
both in the model and in data, and that rescattering slightly increases
temperatures.

About $\phi$ enhancement, our integrated yields per event
without fusion, with fusion, and
with fusion and
rescattering are
3.55, 4.20 and 5.35 respectively, in rough agreement with
experimental data, $7.6 \pm 1.1$ \cite{phi2}.
In Fig. \ref{fig10}
the stopping power is shown, i.e. the $p-\bar p$ rapidity distributions for
central PbPb collisions at SPS, compared with 
the experimental data \cite{stop}, together with the predictions
for RHIC
and LHC energies. This quantity is essentially determined by the string fusion
mechanism and
rescattering only plays a minor role. As discussed for strangeness enhancement,
it has been pointed out
that baryon junction migration \cite{dpmdb,bj1,bj2} will enhance the stopping 
power due to
diagrams additional to the usual ones of the Dual Parton Model.
The inclusion of these diagrams also explains the SPS data. We have not 
taken into account such diagrams to avoid double counting, because in the
fusion of strings they are partially included in an effective way.
In Fig. \ref{fig11} the antiproton rapidity distribution in central PbPb
collisions is presented
and compared to the experimental data \cite{aprot}; a great
suppression of the antiproton yield is seen, due to rescattering.

In Table \ref{tab5}
our results for the ratios between different particles are compared
with the experimental data \cite{ratios} for PbPb central collisions at SPS. 
We observe an overall, rough
agreement with the SPS data, with some excess of $\bar \Lambda$ and
$\bar \Xi^+$, see Fig. \ref{fig7} and comments above.

Let us emphasize
that we obtain a semiquantitative
agreement with the experimental data in PbPb, in three of the
features advocated as signals of QGP production. 
We are only below data in $\Omega$ production by less than a 
factor 2. So we think that our rescattering model, being very simple,
can be useful as a tool to show the trend of such effect and at least
help to tune the initial condition which can be used in transport
models.

Finally,
let us comment on multiplicities in PbPb collisions at SPS energies. For a
centrality of 5 \% (corresponding in the model to $b\leq 3.4$ fm), we get, for
$dN^-/dy$ at $y=0$, 265, 250 and 235 without string fusion, with string fusion,
and with string fusion and rescattering respectively. Experimentally, the NA49
Collaboration gets $196\pm 10$ \cite{na49mult}, while the WA97 Collaboration
gets $178\pm 22$ \cite{wa97mult}. In view of these data the code overestimates
multiplicities. On the other hand,
if we compare the charged
multiplicity per participant (wounded) nucleon
and pseudorapidity unit at midrapidity
versus the number of
wounded nucleons in PbPb collisions at SPS, with data from the WA98
Collaboration \cite{wa98cent}, the trend of data seems to be reproduced,
while their magnitude is underestimated \cite{proximo}.
In
Fig. \ref{fig12} we show the rapidity distribution of negatives compared
with NA49
data \cite{na49mult}.

\section{Predictions for RHIC and LHC}
\label{pred}

Predictions for pseudorapidity and transverse momentum distributions
of charged particles in nucleon-nucleon and central nucleus-nucleus
collisions at RHIC ($\sqrt{s}=200$ GeV per nucleon)
and LHC ($\sqrt{s}=5.5$ and 14 TeV per nucleon for nucleus-nucleus and
nucleon-nucleon collisions respectively)
can be seen\footnote{About 
the reliability of predictions
for RHIC and LHC, see comments in the last paragraph
of Subsection \protect{\ref{strfus}} and in the first paragraph of
Section \protect{\ref{rescatt}}.}
for nucleus-nucleus and nucleon-nucleon collisions respectively)
in Figs. \ref{fig13}, \ref{fig14}
and \ref{fig15}.
While at SPS the influence of string fusion on multiplicities at
midrapidity is of the order $10\div 15$ \%, at RHIC it reaches a $30 \div 35$
\%. In these Figures, the large influence of the hard contribution at
LHC can be observed. Again, the striking fact of the small influence of
parton densities appears, both in nucleon and in nuclear collisions. On
the other hand, the scaling of nucleon-nucleon with the number of
wounded nucleons (the Wounded Nucleon Model \cite{wnm}) gives
predictions which lie far below any of those of our model.

In Fig. \ref{fig9} we plot the inverse exponential slopes of the $m_{T}$ 
distributions for different particles at RHIC.
We see that, compared to the SPS situation, temperatures get higher in
all cases, as expected.

We present our predictions for different particle ratios at
RHIC and LHC in Table \ref{tab6}.
It can be observed that our results are not very 
different
to those of statistical models \cite{lastcall,bm,stat1,stat2}. However,
strangeness enhancement in our case has nothing to 
do with thermal and/or chemical equilibrium.
The main difference
in the predictions for RHIC and LHC between the String Fusion Model
and statistical models
is the overall charged multiplicity, which is respectively 
950 and 3100 for SFM and 1500 and 7600 for
statistical models \cite{noso} (assuming initial temperatures of 500 and
1000 MeV for RHIC and LHC respectively).
Besides, predictions for the stopping power at RHIC and LHC energies are
presented in Fig. \ref{fig10}. Now, a pronounced dip appears at
midrapidities.

Detailed discussions on first RHIC results will be given elsewhere
\cite{proximo}.
Here we simply compare our results with some preliminary data of
the PHOBOS \cite{phobos} and PHENIX \cite{phenix} Collaborations
at RHIC. For charged particles we 
obtain $dN/d\eta \mid_{\mid \eta \mid < 1} = 520$ and 585
for the 6 \% more central AuAu collisions at $\sqrt{s} = 56$ and
130 GeV per nucleon respectively,
to be compared with $408 \pm 12\ {\rm (stat.)} \pm 30\ {\rm (syst.)}$
and $555 \pm 12\ {\rm (stat.)} \pm 35\ {\rm (syst.)}$ ($609 \pm 1\ {\rm
(stat.)} \pm 37\ {\rm (syst.)}$) in PHOBOS (PHENIX). Our prediction for
$\sqrt{s} = 200$ GeV per nucleon with the same centrality cut is
$dN/d\eta \mid_{\mid \eta \mid < 1} = 635$.

\section{Conclusions}
\label{concl}

A Monte Carlo model\footnote{The
code, called psm-1.0,
has been written in Fortran 77 and can be taken as an uuencoded file,
containing instructions
for installation and
use, by anonymous ftp from
ftp://ftp.uco.es/pub/fa1arpen/, or from the web sites
http://www.uco.es/$\,\tilde{ }\,$fa1arpen/ or
http://fpaxp1.usc.es/phenom/, or requested from the authors.}
for nucleon and nuclear collisions in the energy
range going from SPS to LHC has been presented. It is based on a
partonic realization of Regge-Gribov and Glauber-Gribov models and its
translation to strings following the DPM/QGSM ideas. A hard component is
included to reproduce the high transverse momentum tail of the
spectrum. Collectivity is included considering the possibility of fusion
of pairs of strings. Strings are decayed in a conventional way. In order
to tune the parameters of the model and apply it to collisions between
nuclei, a naif model of rescattering has been introduced.

The results of the models turn out to agree reasonably with total
multiplicities, and longitudinal and transverse momentum spectra in the
energy range from SPS to TeVatron. The
agreement with strangeness production, temperature behavior and stopping
at SPS is semiquantitative.

There exist other Monte Carlo models for multiparticle production in
nuclear collisions at ultrarelativistic energies (see \cite{noso} for a
review): RQMD \cite{rqmd},
UrQMD \cite{urqmd},
HIJING \cite{hijing}, DPMJET \cite{dpmjet}, HSD \cite{hsd}, NEXUS
\cite{nexus}, VNI \cite{vni}, AMPT \cite{ampt}, LUCIFER \cite{lucifer},$\dots$
Let us examine the
main similarities and differences, concerning the stage before
rescattering is applied. Both DPMJET and our
model are realizations of the DPM/QGSM which include a hard component,
but we introduce string fusion, while DPMJET considers diquark breaking
diagrams. RQMD takes into account string fusion (and now UrQMD and HIJING
\cite{vance} in a simple way), but no hard part is included either in
RQMD or in UrQMD. The main difference with HIJING lies in the soft
component, which is considered energy independent is HIJING (and in this
way, the multiplicity increase with increasing energy is mainly due to
the hard component), while in our case it increases as an unitarized
supercritical pomeron. VNI is a parton cascade code, in which
the initial stage is mainly generated by hard collisions, with no
hadronic degrees of freedom (strings). AMPT is a hybrid code, which uses
HIJING as initial condition for parton cascade and, after hadronization,
performs hadronic transport. HSD is focused in the transport
of hadronic degrees of freedom, the initial stage not coming from strings
stretched between partons of projectile and target, but considering
strings as
excitations of nucleons in the projectile and target, as in Fritiof
\cite{fritiof}; similar comments can be made for LUCIFER.
Finally, NEXUS is
based in DPM/QGSM, trying to solve the problem of
energy-momentum conservation for both cross sections and
multiparticle amplitudes
at the same time. In our model, energy conservation is strictly
taken into account only for multiparticle amplitudes. Besides NEXUS takes into
account triple pomeron diagrams, which in our case are effectively
included in string fusion.

A detailed
comparison of results of the model with the first RHIC results will be
presented elsewhere \cite{proximo}.
As future developments, strangeness production should be reconsidered
and production of heavier flavors included.
Also fusion of more than two strings and the possibility of
a phase transition like percolation of strings is needed in order to
improve predictions for LHC and study the possibility of
QGP formation in the
framework of string models.

\vskip 1cm

\noi {\bf Acknowledgments:}
We thank M. A. Braun and E. G. Ferreiro, who participated in early
stages
of this work. We also thank G. S. Bali, A. Capella, K. J. Eskola,
A. B. Kaidalov, C. A. Salgado,
Yu. M. Shabelski and K. Werner
for useful discussions, and J. Stachel for comments on
the predictions of the statistical models for RHIC and LHC.
N. A. and C. P.
acknowledge financial support by CICYT of Spain under contract
AEN99-0589-C02 and N. S. A. by Academy of Finland under grant
number 48477.
N. A. and D. S. also thank Universidad de
C\'ordoba and Fundaci\'on Barri\'e de la Maza of Spain respectively, for
financial support.
N. A. thanks Departamento de F\'{\i}sica de
Part\'{\i}culas of the Universidade de Santiago de Compostela, and D. S.
the ALICE Collaboration at CERN, for hospitality during stays in which
part of this work was completed.
Laboratoire de Physique Th\'eorique is Unit\'e Mixte de Recherche --
CNRS
-- UMR n$^{\rm o}$ 8627.

\newpage
\centerline{\bf \large List of figures:}
\vskip 1cm

\noi {\bf 1.} Upper plot: Results in the model for the total (solid
line),
production (dashed line) and hard (dotted line) cross sections
versus $\sqrt{s}$,
compared with
experimental data for total cross sections in pp (filled circles)
and $\bar {\rm p}$p (open circles) collisions taken from
\protect{\cite{pdg}}.
Lower plot: $p_{T min}$ (solid line) used in the model, and model
results for the mean number of total (dashed line) and hard (dotted
line) inelastic collisions per event versus $\sqrt{s}$ computed for the
same collisions as in the upper plot. The ordinary reggeon contribution,
which decreases quickly
with increasing energy, is not included, see text.

\vskip 0.5cm
\noi {\bf 2.} Results of the code for the rapidity distribution
(upper plot), and the $p_T$ distribution for particles with
$2<y_{lab}<4$ (lower plot), of negative
particles in minimum bias pp collisions at $p_{lab}=200$ GeV/c, compared
with
experimental data \protect{\cite{datamarzo,ptpp}}.

\vskip 0.5cm
\noi {\bf 3.} Results of the code for the pseudorapidity distribution
(upper plot), and the $p_T$ distribution for particles with
$|\eta|<2.5$ (lower plot), of charged
particles in minimum bias $\bar {\rm p}$p collisions at $\sqrt{s}=200$
GeV, compared
with
experimental data \protect{\cite{dndetappb200,ptppb200}}. Solid lines
are the results with GRV 94 LO parton densities \protect{\cite{grv94}},
dashed lines with CTEQ5L \protect{\cite{cteq5}}, dotted lines with GRV
98 HO \protect{\cite{grv98}} and dashed-dotted lines
results without semihard contribution. In the $p_T$
distribution, model results have been normalized to experimental data.

\begin{figure}[htb]
\begin{center}
\epsfig{file=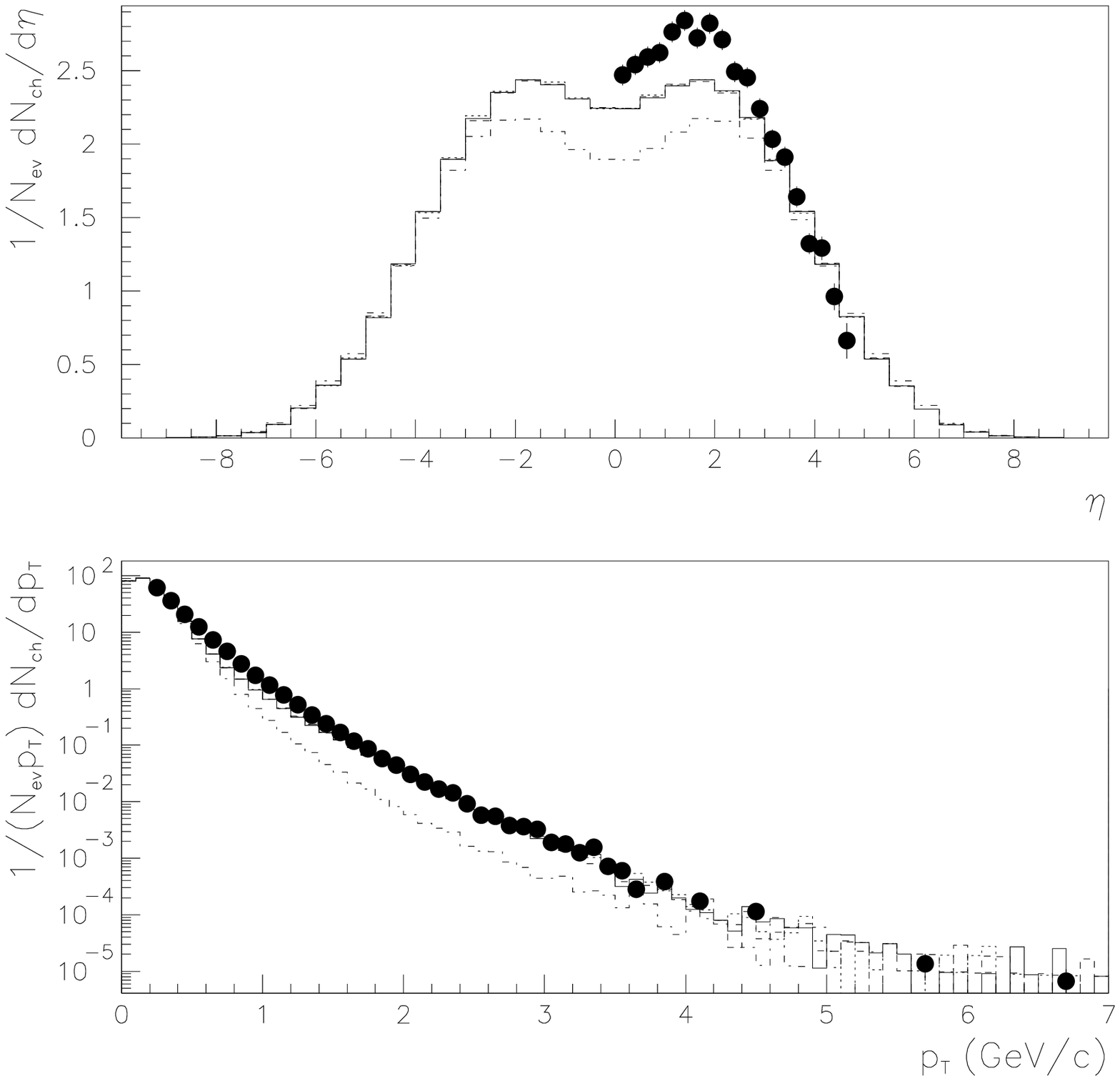,width=15.5cm}
\end{center}
\vskip -1.0cm
\caption{}
\label{fig3}
\end{figure}

\begin{figure}[htb]
\begin{center}
\epsfig{file=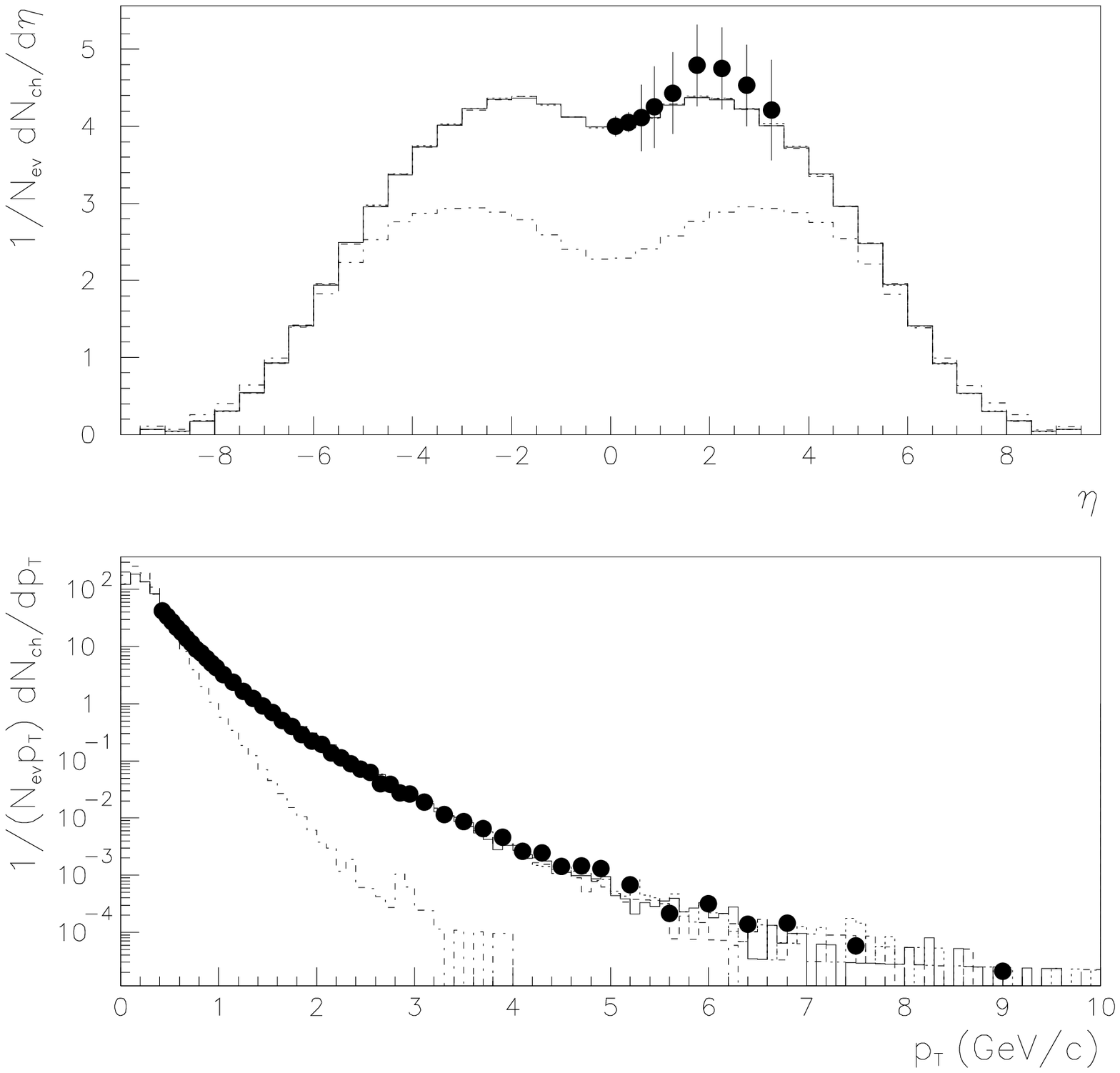,width=15.5cm}
\end{center}
\vskip -1.0cm
\caption{}
\label{fig4}
\end{figure}

\begin{figure}[htb]
\begin{center}
\epsfig{file=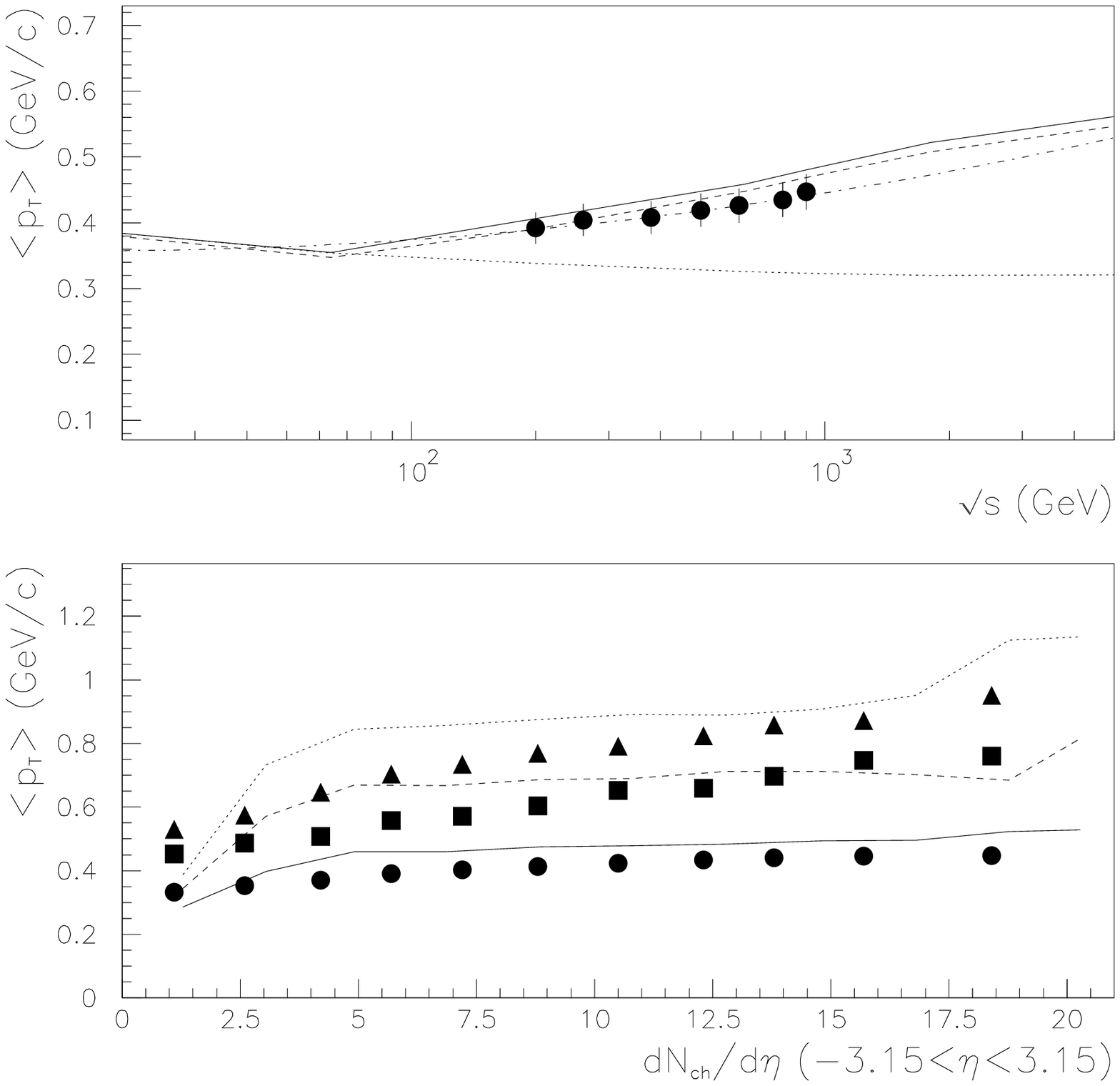,width=15.5cm}
\end{center}
\vskip -1.0cm
\caption{}
\label{fig5}
\end{figure}

\begin{figure}[htb]
\begin{center}
\epsfig{file=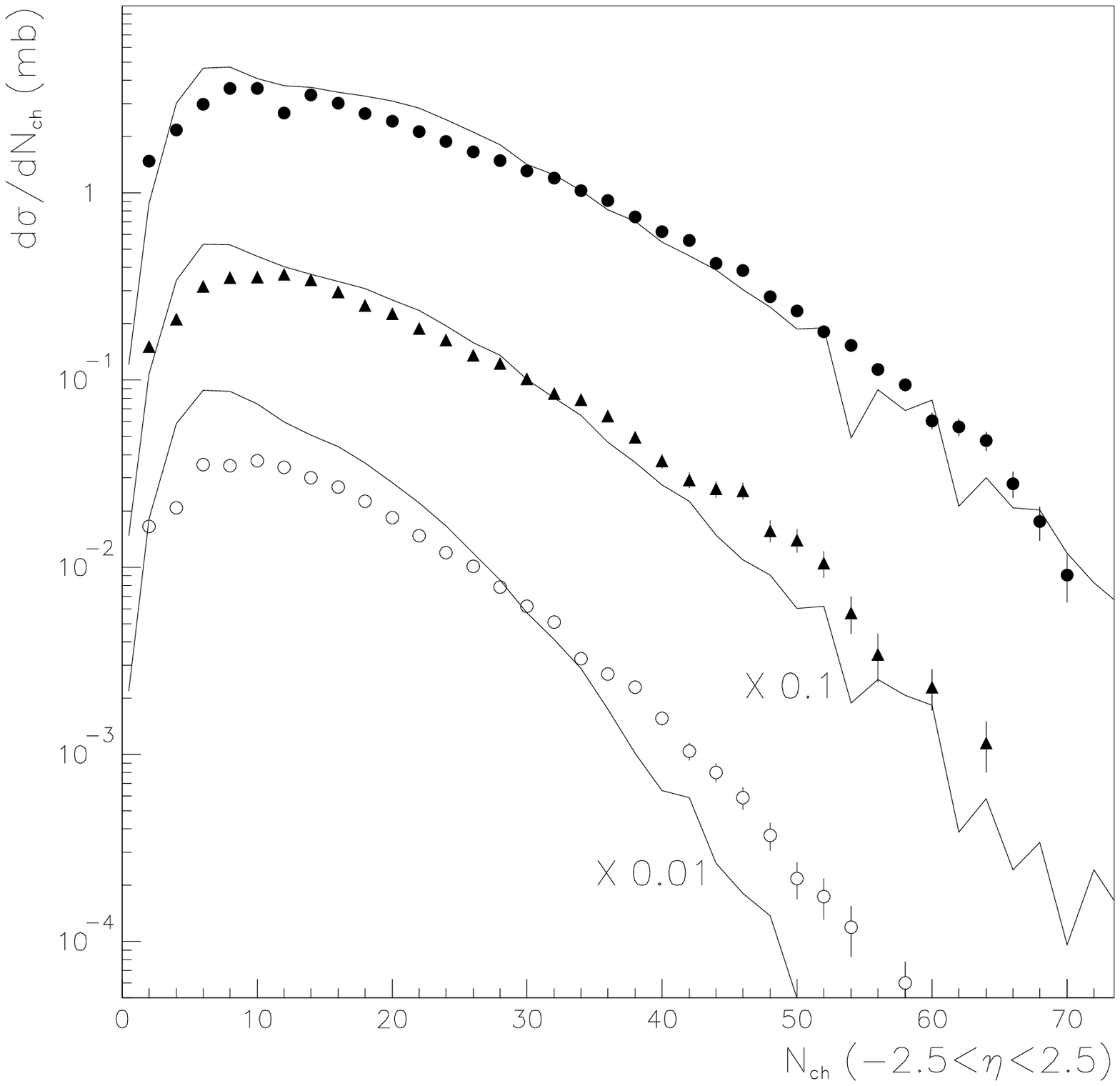,width=15.5cm}
\end{center}
\vskip -1.0cm
\caption{}
\label{fig6}
\end{figure}

\begin{figure}[htb]
\begin{center}
\epsfig{file=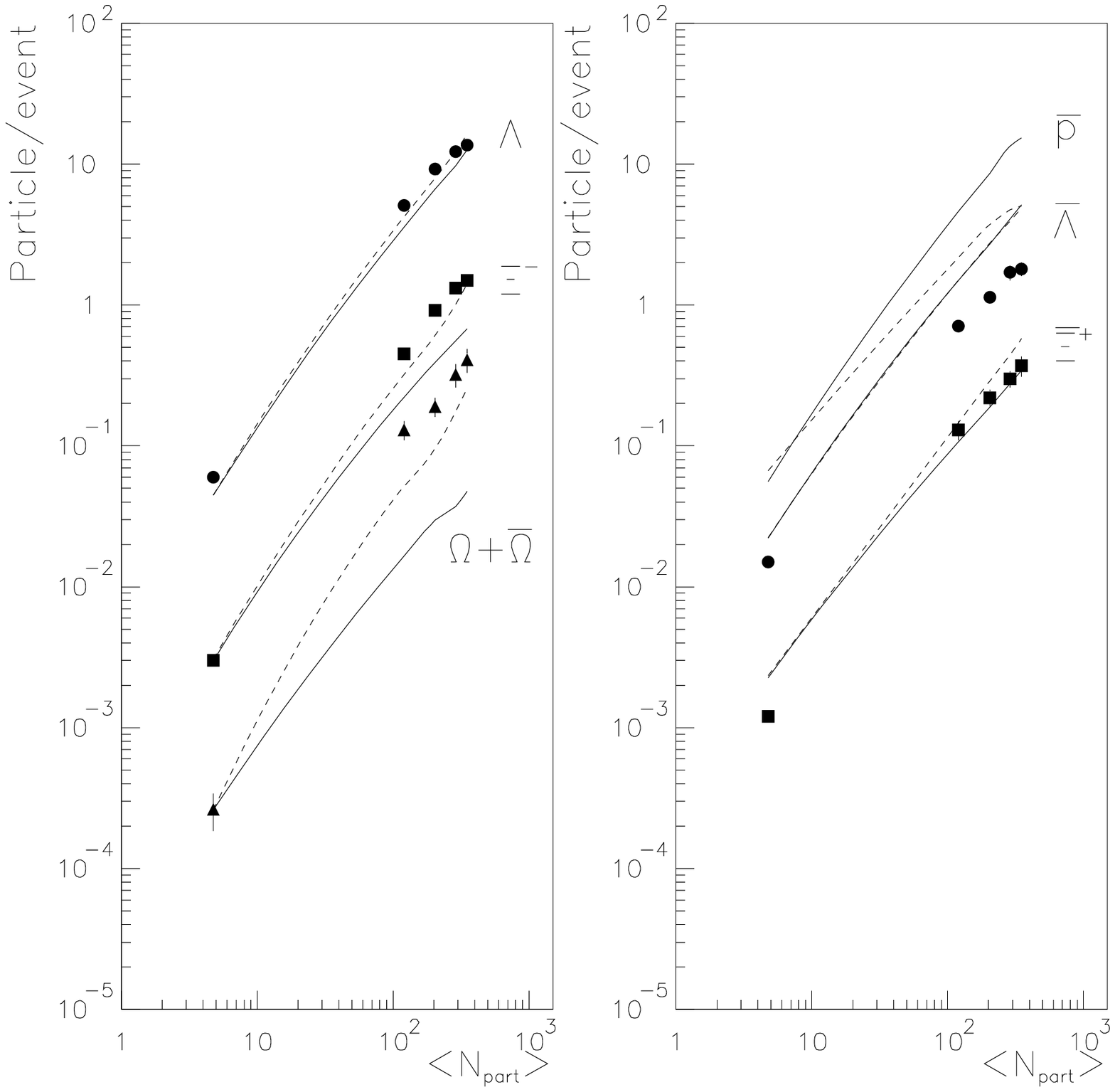,width=15.5cm}
\end{center}
\vskip -1.0cm
\caption{}
\label{fig7}
\end{figure}

\begin{figure}[htb]
\begin{center}
\epsfig{file=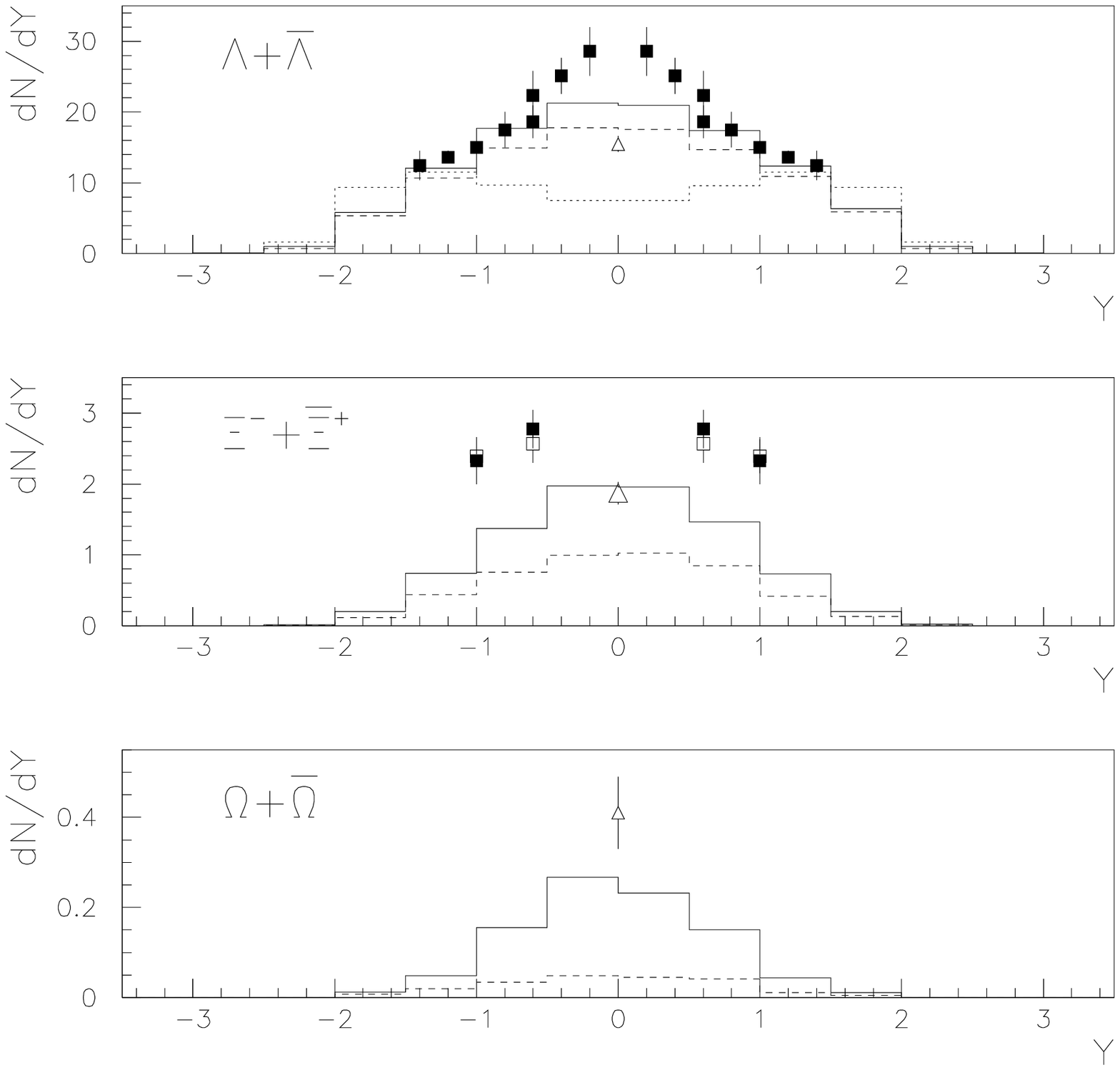,width=15.5cm}
\end{center}
\vskip -1.0cm
\caption{}
\label{fig8}
\end{figure}

\begin{figure}[htb]
\begin{center}
\epsfig{file=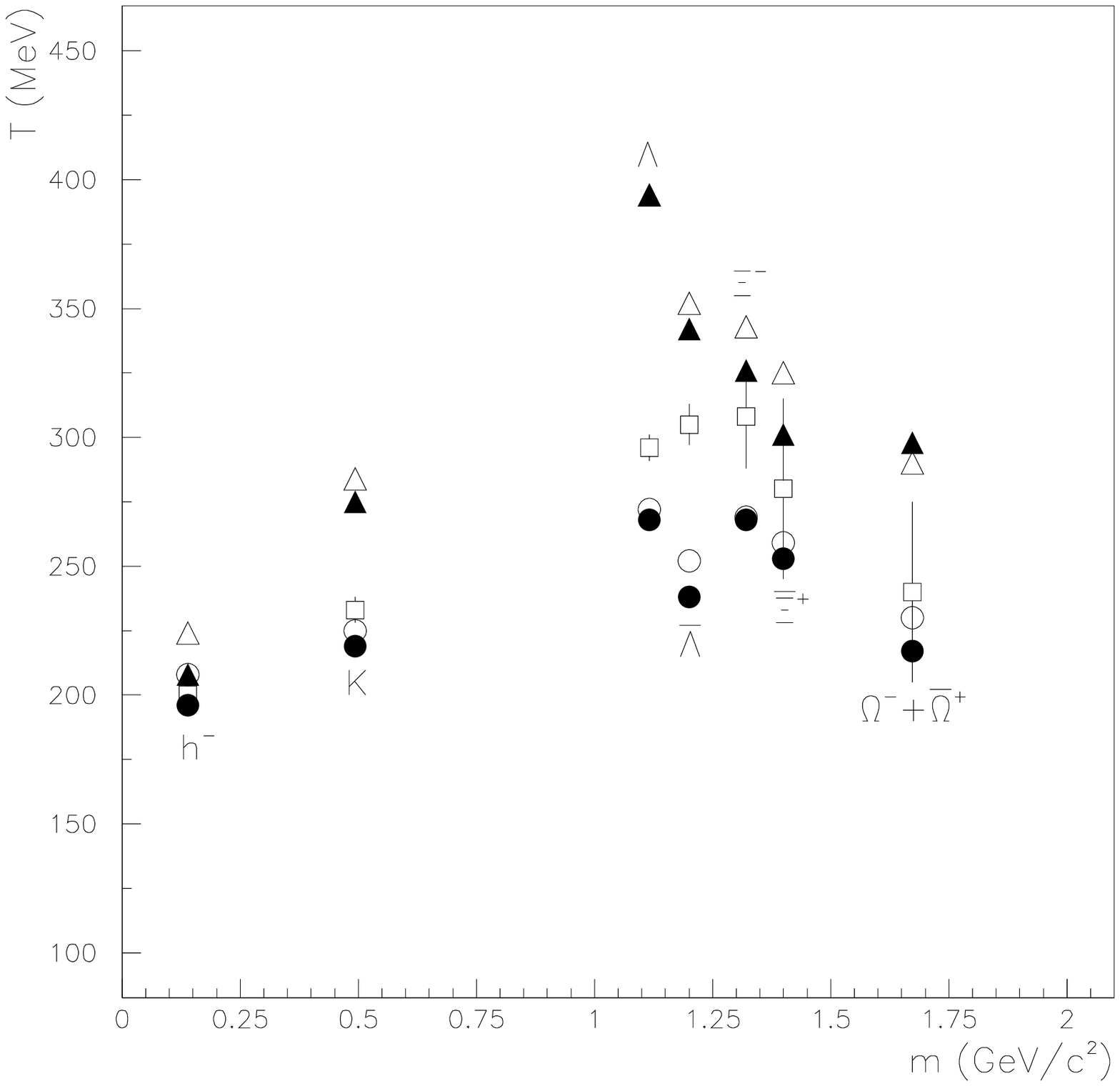,width=15.5cm}
\end{center}
\vskip -1.0cm
\caption{}
\label{fig9}
\end{figure}

\begin{figure}[htb]
\begin{center}
\epsfig{file=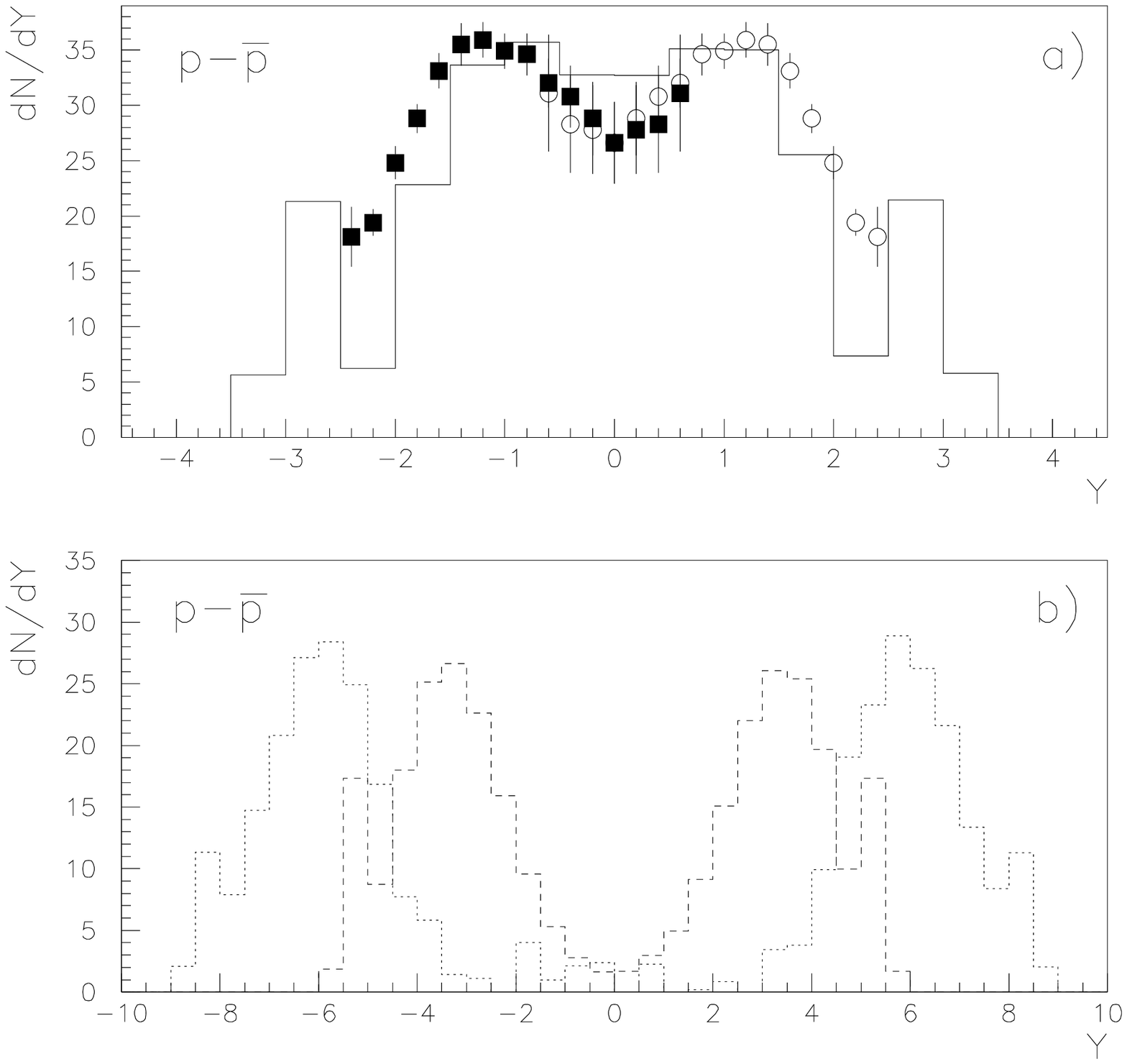,width=15.5cm}
\end{center}
\vskip -1.0cm
\caption{}
\label{fig10}
\end{figure}

\begin{figure}[htb]
\begin{center}
\epsfig{file=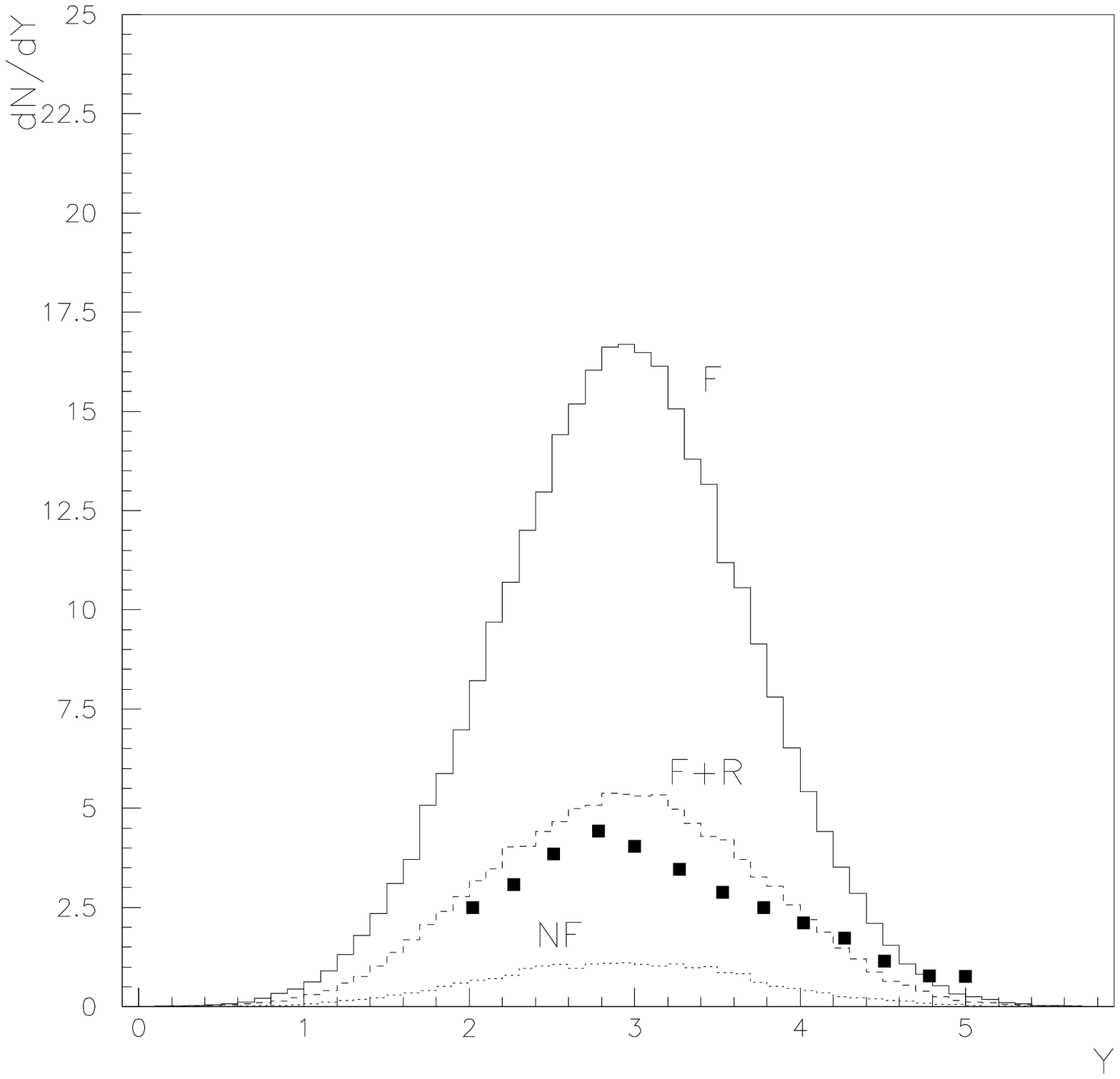,width=15.5cm}
\end{center}
\vskip -1.0cm
\caption{}
\label{fig11}
\end{figure}

\begin{figure}[htb]
\begin{center}
\epsfig{file=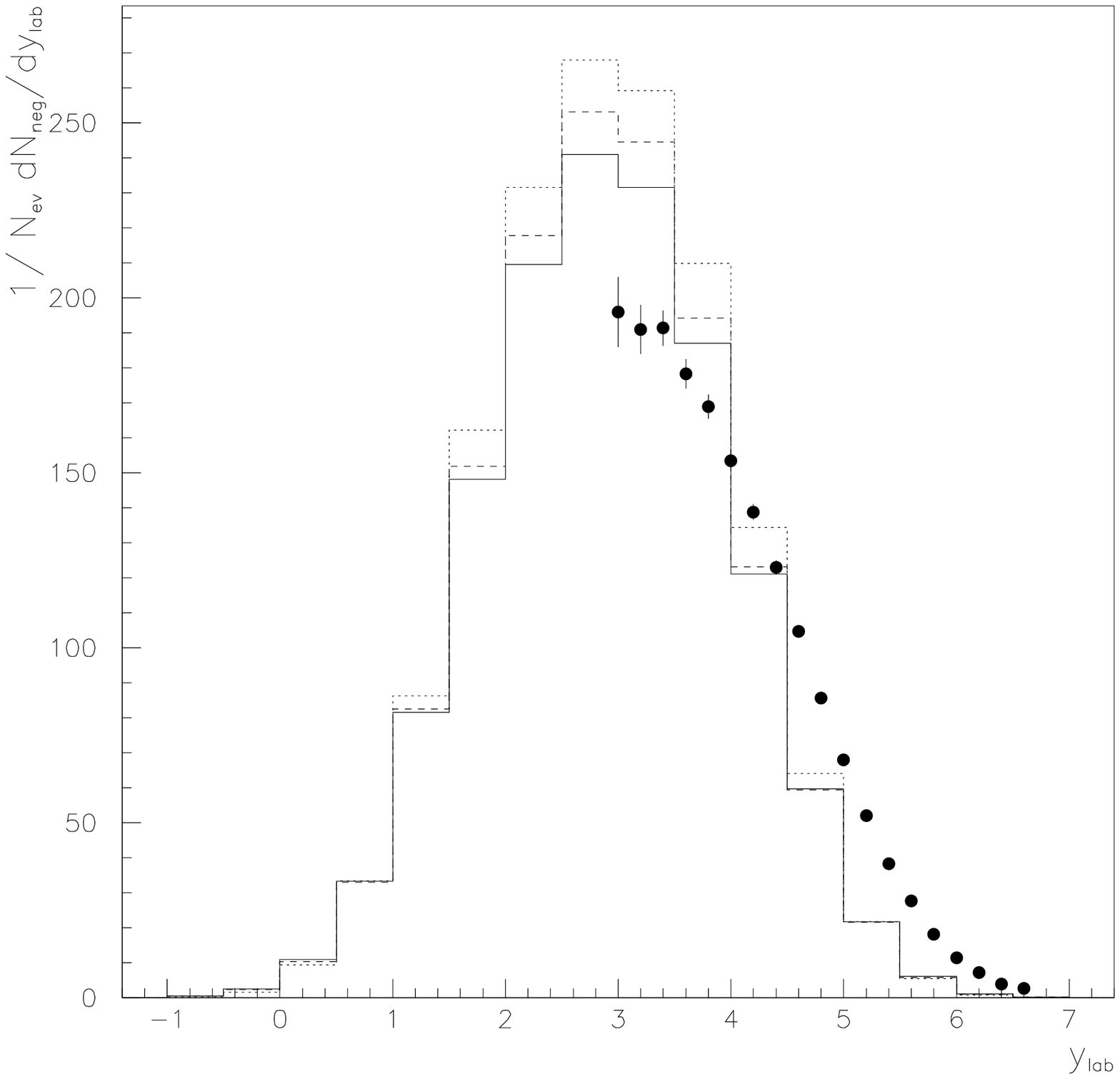,width=15.5cm}
\end{center}
\vskip -1.0cm
\caption{}
\label{fig12}
\end{figure}

\vskip 0.5cm
\noi {\bf 4.} Results of the code for the pseudorapidity distribution
(upper plot), and the $p_T$ distribution for particles with
$|\eta|<1$ (lower plot), of charged
particles in minimum bias $\bar {\rm p}$p collisions at $\sqrt{s}=1.8$
TeV, compared
with
experimental data \protect{\cite{dndetappb1800,ptppb1800}}. Line
convention is the same as in Fig. \protect{\ref{fig3}}. In the $p_T$
distribution, model results have been normalized to experimental data.

\vskip 0.5cm
\noi {\bf 5.} Upper plot: results of the code without hard part (dotted
line), without string fusion (dashed line) and with string fusion
(solid line) for $\langle p_T \rangle$ of charged particles with
$|\eta|<0.5$ in
$\bar {\rm p}$p collisions, versus $\sqrt{s}$, compared with UA1 data
\protect{\cite{ptppb200}} and a parametrization given in this reference
(dashed-dotted line). Lower plot: results of the code for $\langle p_T
\rangle$ of $\pi^\pm$ (solid line), $K^\pm$ (dashed line) and $\bar{\rm
p}$ (dotted line) in $\bar {\rm p}$p collisions at $\sqrt{s}=1.8$ TeV,
versus the pseudorapidity density of charged particles for
$|\eta|<3.15$,
compared
with E735 data \protect{\cite{ptpart}} for $\pi^\pm$ (circles), $K^\pm$
(squares) and
$\bar{\rm
p}$ (triangles).

\vskip 0.5cm
\noi {\bf 6.} Results in the model (solid lines, arbitrarily normalized)
for topological cross sections of charged
particles with $|\eta|<2.5$ in $\bar {\rm p}$p collisions,
compared with experimental data from
\protect{\cite{ptppb200}}. Upper curves and data correspond to
$\sqrt{s}=0.9$ TeV, those in the middle (multiplied by 0.1)
to $\sqrt{s}=0.5$ TeV, and lower curves and data (multiplied by 0.01)
to $\sqrt{s}=0.2$ TeV.

\vskip 0.5cm
\noi {\bf 7.} Yields per unity of rapidity at central rapidity, as a
function
of the number
of wounded nucleons, for $\Lambda$, $\Xi^{-}$ and $\Omega^- +
\bar\Omega^+$
(left),
and for $\bar p$, $\bar\Lambda$ and $\bar\Xi^{+}$ (right), for pPb
collisions and four different centralities in PbPb collisions at SPS
energies. Full lines
represent our
calculation with string fusion, and dashed lines with fusion and
rescattering.
Experimental data are from the WA97 Collaboration
\cite{wa97}.

\vskip 0.5cm
\noi {\bf 8.} Results in the model (dotted line: without fusion, dashed
line: with
fusion, solid line: with fusion and rescattering)
for strange baryon production in central PbPb
collisions (5 \% centrality) at SPS compared with experimental data
from the WA97 Collaboration
\cite{wa97}
(triangles) and the NA49 Collaboration
\cite{na49}
(squares).

\vskip 0.5cm
\noi {\bf 9.} Results in the model (filled circles: with fusion, open
circles: with
fusion and rescattering) for the inverse exponential
slope of the $m_{T}$ distributions at midrapidity
of different particles
versus the mass of the
particles in central (5 \% centrality) PbPb collisions at SPS,
compared with the experimental data of the WA97 Collaboration
\cite{pt}
(3.5 \% centrality, open squares). We also present our predictions for
the same collisions at
RHIC energy with fusion, filled triangles, and with fusion and
rescattering,
open triangles.

\vskip 0.5cm
\noi {\bf 10.} Results in the model
for the $p-\bar p$ rapidity distribution in central
(5 \% centrality) PbPb collisions at SPS (a), solid line), and RHIC
and LHC (b)), dashed and dotted lines respectively),
compared with experimental data at SPS
\cite{stop}.

\vskip 0.5cm
\noi {\bf 11.} Results in the model (dotted line: without fusion,
solid line: with fusion, dashed line: with fusion
and rescattering) for the $\bar p$ rapidity distribution in central
(5 \% centrality) PbPb collisions at SPS, compared with experimental
data
\cite{aprot}.

\vskip 0.5cm
\noi {\bf 12.} Results in the model for the rapidity distribution of
negative
particles in central (5 \% centrality)
PbPb collisions at $\sqrt{s}=17.3$ GeV per nucleon, without fusion
(dotted line), with fusion (dashed line) and with fusion and
rescattering
(solid line), compared with data from NA49 \protect{\cite{na49mult}}.

\vskip 0.5cm
\noi {\bf 13.} Results of the code for the pseudorapidity distributions
(upper plots), and the $p_T$ distributions for particles with
$|\eta|<2.5$ (lower plots), of charged
particles in central ($b\leq 3.2$ fm) AuAu collisions at $\sqrt{s}=200$
GeV per nucleon. In the plots on the left, solid lines are results with
EKS98
\protect{\cite{eks}} parametrization of parton densities inside nuclei,
dashed lines with a parametrization as $F_{2A}$ \protect{\cite{eqc}},
and dotted
lines without modification of parton densities inside nuclei.
In the plots on the right, solid
lines are
results without string fusion, dashed lines with string fusion,
dotted lines are nucleon-nucleon results at the same energy,
scaled by the number of wounded
nucleons (344.6/2), and dashed-dotted lines are results with string
fusion and
rescattering.

\vskip 0.5cm
\noi {\bf 14.} Results of the code for the pseudorapidity distribution
(upper plot), and the $p_T$ distribution for particles with
$|\eta|<2.5$ (lower plot), of charged
particles in minimum bias pp collisions at $\sqrt{s}=14000$
GeV. Line
convention is the same as in Fig. \protect{\ref{fig3}} (the
dashed-dotted line is absent).

\begin{figure}[htb]
\begin{center}
\epsfig{file=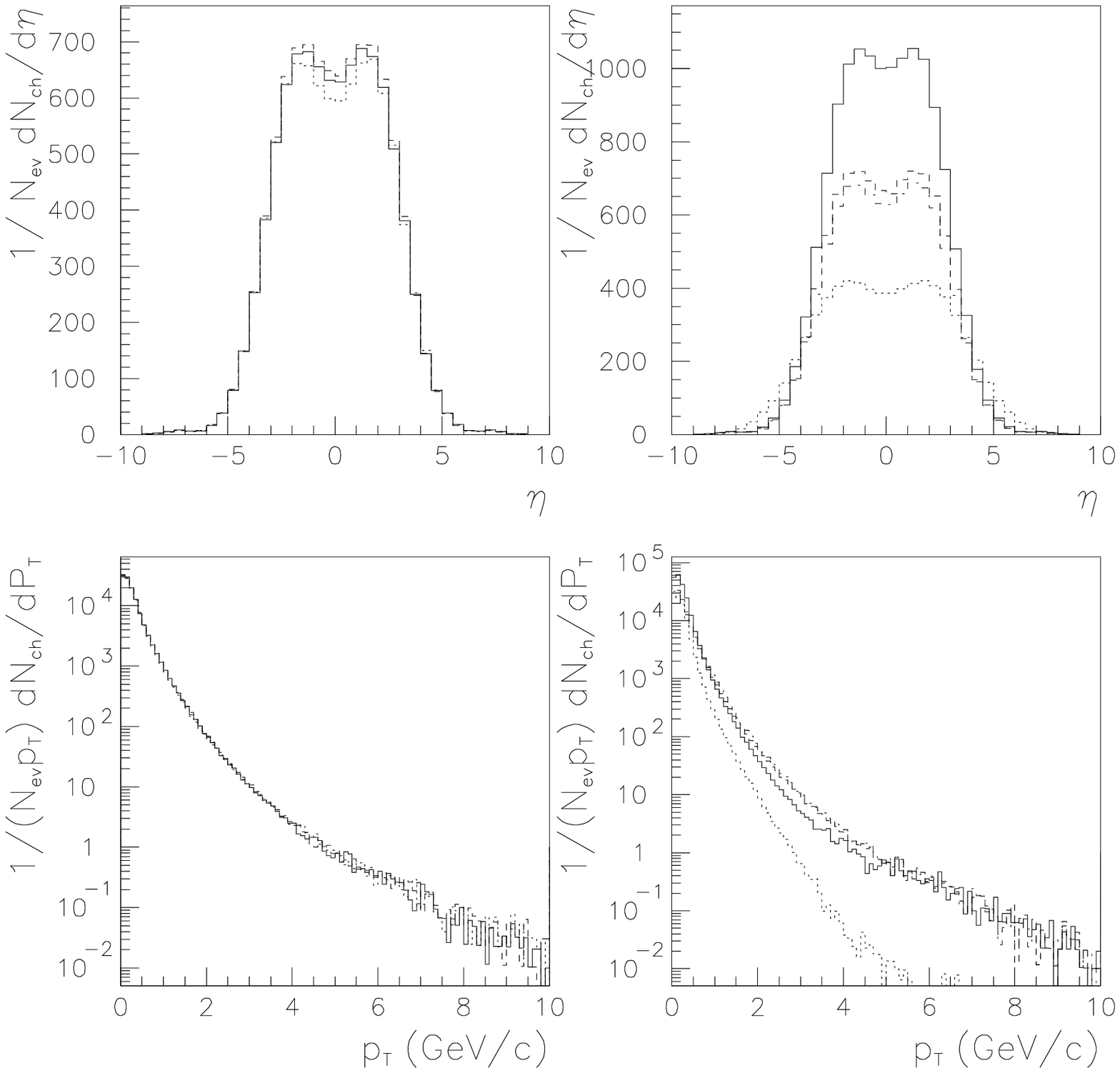,width=15.5cm}
\end{center}
\vskip -1.0cm
\caption{}
\label{fig13}
\end{figure}

\begin{figure}[htb]
\begin{center}
\epsfig{file=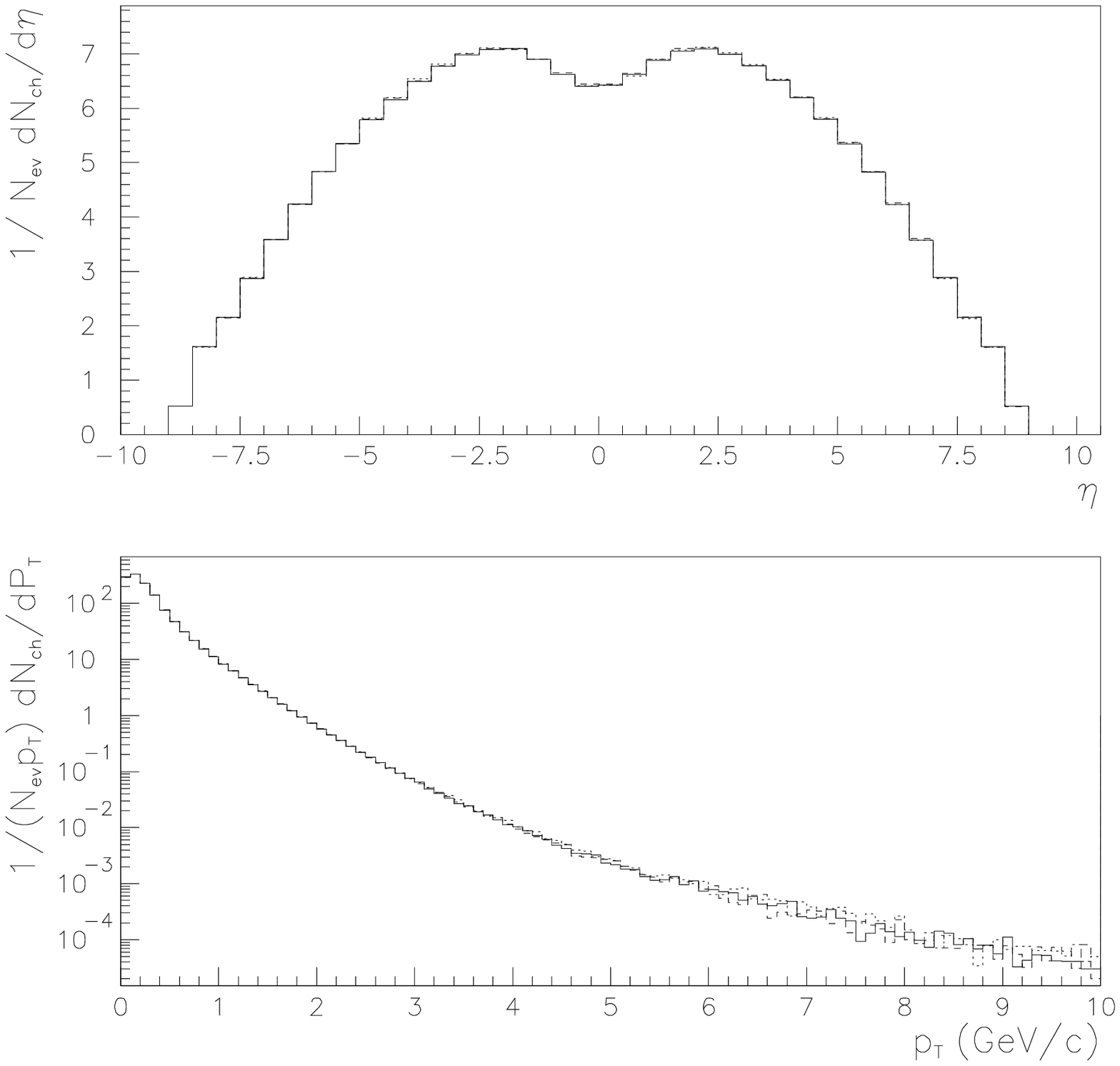,width=15.5cm}
\end{center}
\vskip -1.0cm
\caption{}
\label{fig14}
\end{figure}

\begin{figure}[htb]
\begin{center}
\epsfig{file=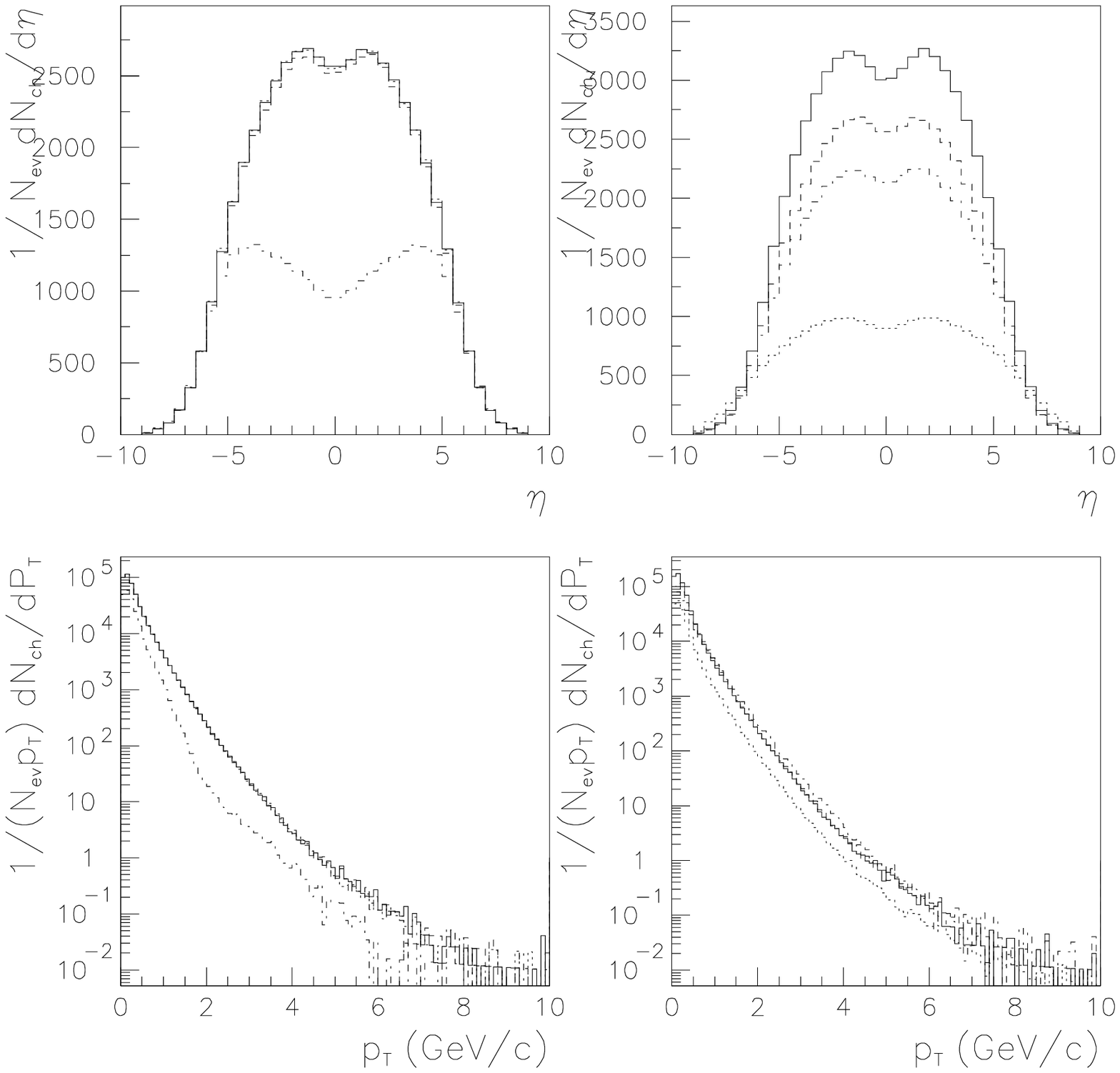,width=15.5cm}
\end{center}
\vskip -1.0cm
\caption{}
\label{fig15}
\end{figure}

\vskip 0.5cm
\noi {\bf 15.} Results of the code for the pseudorapidity distributions
(upper plots), and the $p_T$ distributions for particles with
$|\eta|<2.5$ (lower plots), of charged
particles in central ($b\leq 3.2$ fm) PbPb collisions at $\sqrt{s}=5500$
GeV per nucleon. In the plots on the left, line convention is the same
as in
Fig. \protect{\ref{fig13}} left, but results have been obtained without
rescattering and the
dashed-dotted
lines show results without semihard contribution.
In the plots on the right, solid lines are
results without string fusion, dashed lines with string fusion,
dotted lines are nucleon-nucleon results at the same energy,
scaled by the number of wounded
nucleons (382.1/2), and dashed-dotted lines are results with string
fusion and
rescattering.

\newpage
\centerline{\bf \large List of tables:}
\vskip 1cm

\noi {\bf 1.} Results in the model for mean multiplicities of different
particles in minimum bias pp collisions at $p_{lab}=200$
GeV/c, without and with string fusion, compared with experimental data
\protect{\cite{ppdatasps}}.

\vskip 0.5cm
\noi {\bf 2.} Results in the model for mean multiplicities of different
particles in minimum bias pp collisions at $\sqrt{s}=27.5$
GeV, without and with string fusion, compared with experimental data
\protect{\cite{ppdata27}}.

\vskip 0.5cm
\noi {\bf 3.} Results in the model for mean multiplicities of negative
particles in minimum bias pA collisions at $p_{lab}=200$
GeV/c, without and with string fusion, compared with experimental data
for pS \protect{\cite{dataps}}, pAr and pXe \protect{\cite{datamarzo}},
and pAg and pAu
\protect{\cite{datapag}}.

\vskip 0.5cm
\noi {\bf 4.} Results in the model for mean multiplicities of different
particles in central ($b\leq 1.3$ fm) SS  collisions at $\sqrt{s}=19.4$
GeV per nucleon,
compared with experimental data
\protect{\cite{datass}}. Results are presented without fusion (NF),
with fusion (F), and with fusion and rescattering (FR).

\vskip 0.5cm
\noi {\bf 5.} Results in the model for different particle ratios at
midrapidity
in
central (30 \% centrality)
PbPb collisions at SPS, compared with experimental data
\protect{\cite{ratios}}, following the same convention as in Table
\protect{\ref{tab4}}.

\vskip 0.5cm
\noi {\bf 6.} Results in the model for different particle ratios at
midrapidity
in
central ($b\leq 3.2$ fm) AuAu collisions at RHIC and PbPb collisions at
LHC,
following the same convention as in Table \protect{\ref{tab4}}.
For comparison, results from other models (Quark Coalescence Model (QCM)
\protect{\cite{lastcall}},
Rafelski
\protect{\cite{lastcall}}
and B-M
\protect{\cite{lastcall,bm}})
for RHIC are
included.

\newpage
\centerline{\bf \large Figures:}

\newpage

\newpage

\newpage

\newpage

\newpage

\newpage

\newpage

\newpage

\newpage

\newpage

\newpage

\newpage

\newpage

\newpage

\newpage
\centerline{\bf \large Tables:}

\vskip 1cm

\begin{table}[htb]
\begin{center}
\begin{tabular}{cccc}
\hline\hline
 & No fusion & Fusion & Experiment \\
\hline \hline
charged & 7.89  & 7.81 &  $7.69 \pm 0.06$ \\ \hline
negatives & 2.95  & 2.90 &   $2.85 \pm 0.03$ \\ \hline
p & 1.18 &  1.19 &  $1.34 \pm 0.15$ \\ \hline
$\pi^+$  &  3.40 &  3.33 & $3.22  \pm 0.12$ \\ \hline
$\pi^-$  & 2.69  &  2.63 & $2.62 \pm 0.06$ \\ \hline
$\pi^0$  &  3.73  &  3.68 & $3.34 \pm 0.24$ \\ \hline
$K^+$  &  0.31  &  0.32  & $0.28\pm 0.06$ \\ \hline
$K^-$  &  0.18 &   0.18  & $0.18 \pm 0.05$ \\ \hline
$\Lambda$ & 0.223 &    0.231 & $0.096 \pm 0.010$ \\ \hline
$\bar \Lambda$ & 0.029 & 0.033 & $0.0136 \pm 0.0040$ \\ \hline
$\bar {\rm p}$ &  0.059 & 0.070 & $0.05 \pm 0.02$ \\ \hline \hline
\end{tabular}
\end{center}
\caption{}
\label{tab1}
\end{table}

\begin{table}[htb]
\begin{center}
\begin{tabular}{cccc}
\hline\hline
 & No fusion & Fusion & Experiment \\
\hline\hline
p & 1.20 & 1.21 & $1.20 \pm 0.12$ \\ \hline
$\pi^+$  & 4.04 & 3.94 & $4.10 \pm 0.26$ \\ \hline
$\pi^-$  & 3.32 & 3.23 & $3.34 \pm 0.20$ \\ \hline
$\pi^0$  & 4.47 & 4.38 & $3.87 \pm 0.28$ \\ \hline
$K^+$  &0.38 & 0.38 & $0.33 \pm 0.02$ \\ \hline
$K^-$  &0.25 & 0.24 & $0.22 \pm 0.01$ \\ \hline
$\Lambda$ & 0.245 & 0.251 & $0.13 \pm 0.01$ \\ \hline
$\bar \Lambda$ & 0.045 & 0.049 & $0.020 \pm 0.005$ \\ \hline
$\bar {\rm p}$ & 0.088 & 0.100 & $0.063 \pm 0.002$ \\ \hline \hline
\end{tabular}
\end{center}
\caption{}
\label{tab2}
\end{table}

\begin{table}[htb]
\begin{center}
\begin{tabular}{cccc}
\hline\hline
 & No fusion & Fusion & Experiment \\
\hline\hline
pS & 5.01 & 4.86 & $5.10\pm 0.20$ \\ \hline
pAr & 5.31 & 5.12 & $5.39\pm 0.17$ \\ \hline
pAg & 6.57 & 6.28 & $6.2\pm 0.2$ \\ \hline
pXe & 6.89 & 6.56 & $6.84\pm 0.13$ \\ \hline
pAu & 7.54 & 7.16 & $7.0\pm 0.4$ \\ \hline \hline
\end{tabular}
\end{center}
\caption{}
\label{tab3}
\end{table}

\begin{table}[htb]
\begin{center}
\begin{tabular}{ccccc}
\hline\hline
 & NF & F & FR & Experiment \\
\hline\hline
negatives & 108.2 & 101.3 & 100.7 & $98\pm 3$ \\ \hline
$K^+$ & 9.7 & 10.4 & 10.8 & $12.5\pm 0.4$ \\ \hline
$K^-$ & 7.1 & 7.2 & 7.4 & $6.9\pm 0.4$ \\ \hline
$\Lambda$ & 5.0 & 5.9 & 6.0 & $9.4 \pm 1.0$ \\ \hline
$\bar \Lambda$ & 0.4 & 1.1 & 1.2 & $2.2 \pm 0.4$ \\ \hline
$\bar {\rm p}$ & 0.82 & 3.23 & 2.80 & \\ \hline
$\Xi^-$ & 0.024 & 0.186 & 0.205 & \\ \hline
$\bar \Xi^+$ & 0.028 & 0.097 & 0.102 & \\ \hline
$\Omega^-$ & 0.001 & 0.007 & 0.010 & \\ \hline
$\bar \Omega^+$ & 0.001 & 0.005 & 0.007 & \\ \hline \hline
\end{tabular}
\end{center}
\caption{}
\label{tab4}
\end{table}

\begin{table}[htb]
\begin{center}
\begin{tabular}{ccccc}
\hline
\hline
& NF & F & FR & Experiment\\
\hline
\hline
$\bar \Lambda/\Lambda$ & 0.14 & 0.41 & 0.34 & $0.128 \pm 0.012$ \\ \hline
$\bar \Xi^{+}/\Xi^{-}$ & 1.12 & 0.52 & 0.45 & $0.266 \pm 0.028$ \\ \hline
$\bar \Omega^{+}/\Omega^{-}$ & 0.75 & 0.88 & 0.59 & $0.46 \pm 0.15$ \\ \hline
$\Xi^{-}/\Lambda$
& 0.01 & 0.06 &0.08  & $0.093 \pm 0.007$ \\ \hline
$\bar \Xi^{+}/\bar \Lambda$ & 0.08 & 0.07 & 0.10 & $0.195 \pm 0.023$
\\ \hline
$\Omega/\Xi$ & 0.05 & 0.05 & 0.11 & $0.195 \pm 0.028$ \\
\hline
\hline
\end{tabular}
\end{center}
\caption{}
\label{tab5}
\end{table}

\begin{table}[htb]
\begin{center}
\begin{tabular}{cccccccc}
\hline
\hline
& RHIC(F) & RHIC(FR) & QCM & Rafelski & B-M & LHC(F) & LHC(FR)\\
\hline
\hline
$\bar \Lambda/\Lambda$ & 1.01 & 0.90 & 0.69 & $0.49 \pm 0.15$ & 0.91
&1.00 &0.98 \\ \hline
$\bar \Xi^{+}/\Xi^{-}$ & 0.96 & 0.97 & 0.83 & $0.68\pm 0.15$
& 1.0 &0.98 &0.95 \\ \hline
$\bar \Omega^{+}/\Omega^{-}$ & 1.00 & 1.25 & 1.0 & 1.0 & 1.0 &0.76 &1.03
\\ \hline
$\Xi^{-}/\Lambda$ & 0.10 & 0.15 &$-$ &$0.18\pm 0.02$ & 0.13 &0.09 &0.25
\\ \hline
$\bar \Xi^{+}/\bar \Lambda$ & 0.10 & 0.16 & $-$&
$0.25 \pm 0.03$ & 0.14 &0.09 &0.24 \\ \hline
$\Omega/\Xi$ & 0.07 &0.26  &$-$ &$0.14\pm 0.03$ &0.20 & 0.05& 0.40\\ \hline
$\bar \Lambda/\bar p$ & 0.40 &0.71  & $-$ & $2.4 \pm 0.3$ & 0.52
&0.35 &0.82 \\ \hline
$\bar p / p$ & 0.93 & 0.90 & 0.58 &$0.34^{+0.37}_{-0.12}$ & 0.90 &1.00
&1.04 \\
\hline
\hline
\end{tabular}
\end{center}
\caption{}
\label{tab6}
\end{table}

\end{document}